\documentclass[12pt,english]{article}
\usepackage[T1]{fontenc}
\usepackage[utf8]{inputenc}
\usepackage{xcolor}
\usepackage{colortbl}
\usepackage{babel}
\usepackage{float}
\usepackage{url}
\usepackage{amsmath}
\usepackage{amsthm}
\usepackage{amssymb}
\usepackage{geometry}
\usepackage{setspace}
\usepackage[bookmarks=false,
 breaklinks=false,pdfborder={0 0 1},backref=section,colorlinks=true]
 {hyperref}
\hypersetup{
 citecolor=black,linkcolor=black,urlcolor=black}

\begin{document}
\title{Ultrametric organization of energy landscapes on random Erdős--Rényi
graphs: topological origin of barrier hierarchy}
\author{ A.\,P.~Zubarev \\
 \textit{Volga State Transport University,} \\
 \textit{Pervyi Bezyamyannyi pereulok 18, Samara, 443066, Russia;}
\\
 \textit{Samara University,} \\
 \textit{Moskovskoe shosse 34, Samara, 443123 Russia} \\
 e-mail:\thickspace{}\texttt{apzubarev@mail.ru} }
\maketitle
\begin{abstract}
In the present work, we investigate the hypothesis of the ultrametric
organization of abstract energy landscapes defined on random Erdős--Rényi
graphs. Within the proposed model, each vertex of the graph is assigned
a random free energy uniformly distributed on a given interval, and the
kinetics of the system is described by a Markov process with a Kramers
transition rate matrix. Based on the spectral decomposition of this
matrix, a kinetic metric between basins of attraction is constructed.
Computational experiments for graphs with $V=5000$ vertices and $E=5000$
edges (average degree $\langle k\rangle=2.0$) demonstrate that this
metric reveals a high degree of nontrivial ultrametricity, which
monotonically increases from $\approx42\%$ at an energy interval width
of $\Delta F=10$ kJ/mol to $\approx96\%$ at $\Delta F=1000$ kJ/mol.
For a rigorous mathematical justification of this phenomenon, a limit
theorem is formulated and proved, establishing that as the width of
this interval tends to infinity, the logarithmic asymptotics of the
proposed metric converges pointwise to the classical single-linkage
ultrametric. The analysis of computational experiments shows that for
finite values of the interval width, suboptimal paths and thermal
fluctuations introduce corrections that violate the strong triangle
inequality; however, in the regime of large energy fluctuations, their
contribution is exponentially suppressed, and consequently the metric
asymptotically acquires ultrametric properties. The obtained results
indicate that the ultrametric organization of energy landscapes
represents a universal asymptotic property of networks of arbitrary
topology with a strongly rugged relief, which for sparse, locally
tree-like graphs manifests itself already at moderate energy spreads.
\end{abstract}

\section{Introduction}

\label{sec_intr}

The hierarchical organization of energy landscapes is one of the
central concepts of modern physics of disordered systems, spin glass
theory, and biophysics of macromolecules. Mathematically, such
organization is described by ultrametric spaces -- metric spaces in
which for any three points $x$, $y$, $z$ the metric $d$ satisfies the
condition
\begin{equation}
d(x,y)\le\max\{d(x,z),d(z,y)\},
\end{equation}
which is called the strong triangle inequality. The fundamental
properties of ultrametrics, including their connection with rooted
trees and hierarchical clustering, are detailed in the classic review
\cite{rammal1986} and monographs (see, e.g., \cite{Hartigan,Sneath}).
These ideas have found applications in a wide variety of fields -- from
phylogenetic analysis \cite{Lake} and taxonomy theory to statistical
physics and $p$-adic analysis \cite{ALL,ALL_1,DKKM}. A key event that
stimulated interest in ultrametricity in physics was Parisi's discovery
of the ultrametric organization of the state space in the
Sherrington--Kirkpatrick spin glass model \cite{parisi1979,parisi1980}.
Further studies \cite{mezard1984,mezard1987} and rigorous mathematical
proofs \cite{talagrand2003,talagrand2011,panchenko2013} confirmed that
ultrametricity is not an artificial construction but reflects the real
geometry of landscapes with many competing minima separated by barriers
of increasing height. Almost simultaneously with these works,
Frauenfelder and co-workers, investigating the kinetics of CO binding to
myoglobin, formulated the hypothesis of a hierarchical organization of
the conformational space of proteins \cite{frauenfelder1988,frauenfelder1991,frauenfelder2000}.
According to this hypothesis, protein dynamics represents a random walk
on an ultrametric tree of metastable substates grouped into clusters
separated by barriers. The development of these ideas led to the
formation of a whole research direction at the intersection of $p$-adic
analysis and molecular biophysics (see, e.g., \cite{avetisov2002,avetisov2009,ABZ_2014,bikulov2021}),
within which concrete models of $p$-adic diffusion on energy landscapes
of proteins were constructed.

The motivation for the present work came from recent results obtained
by the author when analyzing the ultrametricity of a kinetic metric on
the space of secondary structures of ribonucleic acids (RNA)
\cite{zubarev2026_rna}, as well as subsequent unpublished numerical
studies of protein molecules, which demonstrated that the kinetic metric
on the space of quasi-equilibrium states of biopolymers possesses an
unexpectedly high degree of nontrivial ultrametricity. This observation
gave rise to a fundamental question -- is the discovered hierarchy a
unique consequence of biological evolution optimizing the energy
landscape of a biological molecule, or does it represent a universal
asymptotic property of any networks with a rugged energy relief, which,
in the case of sparse, locally tree-like graphs, manifests itself
especially brightly and is accessible for direct numerical observation?

To answer this question, in this work we apply the mathematical
apparatus developed in \cite{zubarev2026_rna} to a model of a random
energy landscape defined on Erdős--Rényi graphs. We generate an ensemble
of sparse random graphs, assign a random energy to each vertex from a
uniform distribution on a given interval, and apply algorithms for
basin finding, spectral decomposition, and kinetic metric computation.
A principal difference of such a system from biopolymer systems is the
complete absence of physicochemical constraints on the graph topology
and energy distribution, which allows us to investigate the
ultrametricity of the system in its purest, abstract form.

The central result of our work is a mathematical proof of the fact
that, as the width of the energy distribution tends to infinity, the
logarithmic asymptotics of the kinetic metric constructed from the
spectral decomposition of the Kramers transition rate matrix converges
pointwise to the classical single-linkage ultrametric. This limit
theorem not only explains the monotonic growth of the degree of
nontrivial ultrametricity observed in computational experiments with
increasing free energy distribution interval, but also establishes a
direct mathematical connection between two fundamental constructions --
the kinetic metric, which accounts for all possible transition paths,
and the minimax metric (single-linkage metric), determined by a single
optimal path.

The paper is organized as follows. Section \ref{sec_model} describes
the random Erdős--Rényi graph model and phase transitions in such
networks. Section \ref{sec_kramers} is devoted to the construction of
the Kramers transition rate matrix and the proof of its spectral
properties. Section \ref{sec_bas} describes the gradient descent
algorithm for identifying macrostates, the spectral decomposition
procedure, and the construction of the kinetic metric. Section
\ref{sec_ultra} contains the definition of a formal ultrametricity
criterion. Section \ref{sec_results} presents the results of the
computational experiment and their analysis. Section
\ref{sec_limit_theorem} contains the formulation and proof of the limit
theorem on the convergence of the kinetic metric to the single-linkage
metric. Section \ref{sec_discussion} discusses the physical meaning of
the obtained results, provides a comparison with $p$-adic models, and
outlines directions for further research.

\section{Random Energy Landscape Model on Sparse Graphs}

\label{sec_model}

As the basic topological structure, we use the classical Erdős--Rényi
random graph model $G(V,E)$ \cite{erdos1959,frieze2015}. The graph is
constructed on a set of vertices $\mathcal{M}=\{1,2,\dots,V\}$, which
correspond to the Markov microstates of the system. The edges of the
graph $\mathcal{E}$ are formed by randomly and equiprobably selecting
$E$ unordered pairs of vertices without replacement, which guarantees
the absence of multiple edges.

The key parameter determining the global topology of an Erdős--Rényi
graph is the average vertex degree:
\[
\langle k\rangle=\frac{2E}{V}.
\]
Random graph theory establishes the existence of a phase transition at
$\langle k\rangle=1$. Namely, the following statement holds
\cite{frieze2015}. Consider a random graph with $V$ vertices and $E$
edges. Let $V\to\infty$, while the average vertex degree $\langle k\rangle=2E/V$
remains constant. Then for $\langle k\rangle<1$, all connected components
of the graph are almost surely trees or contain exactly one cycle, and
the size of the largest connected component is of order $O(\log V)$.
At $\langle k\rangle=1$, the size of the largest connected component is
of order $O(V^{2/3})$. For $\langle k\rangle>1$, a unique giant
connected component almost surely emerges, containing a positive
fraction $S>0$ of all vertices, where $S$ is determined as the unique
positive solution of the equation $S=1-\exp(-\langle k\rangle S).$
All other connected components have size $O(\log V)$.

The most important property of sparse Erdős--Rényi graphs is their
local tree-likeness. A graph is called locally tree-like if for any
vertex $v$, its $r$-neighborhood $N_{r}(v)=\{u:d_{\mathrm{graph}}(u,v)\le r\}$
with high probability does not contain cycles for $r=O(\log\log V)$.
For an Erdős--Rényi graph with $\langle k\rangle=O(1)$, the probability
of having a cycle of length $l$ in the neighborhood of an arbitrary
vertex is estimated as $O(\langle k\rangle^{l}/V)$, which tends to zero
as $V\to\infty$. This means that locally the graph is isomorphic to a
Cayley tree (regular Bethe tree) with branching degree $\langle k\rangle$.
This property plays an important heuristic role for understanding the
kinetics: the absence of local cycles means that between any two remote
vertices there exists essentially a unique trunk path. Alternative paths
require going to higher hierarchical levels of the tree, which forms a
natural hierarchy of barriers. As will be shown below, although
ultrametricity in the asymptotic limit of infinite energy spread arises
on graphs of arbitrary topology, it is precisely the local
tree-likeness of sparse graphs that ensures the manifestation of this
effect at finite, physically reasonable values of $\Delta F$.

Consider some realization of the Erdős--Rényi graph $G=(\mathcal{M},\mathcal{E})$.
Assign to each vertex $i\in\mathcal{M}$ of the graph a random value of
free energy $F_{i}$, where the vertex energies are chosen independently
from a uniform distribution on a given interval $(F_{\min},F_{\max})$.
We next consider a Markov process of random walk on the graph
$G=(\mathcal{M},\mathcal{E})$. To describe the evolution of the system,
we use the master equation:
\[
\frac{df_{i}(t)}{dt}=\sum_{j:\{i,j\}\in\mathcal{E}}\left(k_{j\to i}f_{j}(t)-k_{i\to j}f_{i}(t)\right),
\]
where $f_{i}(t)$ is the probability of being in microstate (vertex)
$i$, $k_{i\to j}$ are the rate constants for transitions between
microstates, and the summation is carried out only over neighboring
vertices, i.e., vertices connected by edges.

\section{Kramers Transition Rate Matrix and its Spectral Properties}

\label{sec_kramers}

According to the theory of chemical reaction rates \cite{Kramers}, the
transition rate constant is determined by the height of the energy
barrier. In the absence of information about the exact geometry of the
saddle points, a conservative lower bound of the barrier via the free
energies of the initial and final states is used: for two neighboring
microstates $i$ and $j$, the rate constant for the transition $i\to j$
has the form
\begin{equation}
k_{i\to j}=\nu_{0}\exp\left(-\frac{\max\{F_{i},F_{j}\}-F_{i}}{RT}\right),\label{eq_kramers_rate_asymmetric}
\end{equation}
where $\nu_{0}>0$ is the frequency factor (preexponential multiplier)
with dimension $[\text{time}]^{-1}$, $T$ is the temperature,
$R$ is the universal gas constant, $F_{i}$ is the free energy expressed
in molar units, and the quantity $\max\{F_{i},F_{j}\}$ is used as a
lower bound for the free energy of the transition state. This estimate
(\ref{eq_kramers_rate_asymmetric}) is standard for models in which the
exact geometry of the energy surface is unavailable \cite{Zuckerman}.
The constants $k_{i\to j}$ satisfy detailed balance:
\[
k_{i\to j}\cdot w_{i}=k_{j\to i}\cdot w_{j},
\]
where $w_{i}=\exp(-F_{i}/RT)$ is the Boltzmann weight of microstate
$i$.

Define the matrix $K$ of size $V\times V$, where $V=|\mathcal{M}|$,
as follows:
\[
K_{ij}=\begin{cases}
\nu_{0}\exp\left(-\dfrac{\max\{F_{i},F_{j}\}-F_{j}}{RT}\right), & \text{for }\{i,j\}\in\mathcal{E},\\[10pt]
0, & \text{for }\{i,j\}\notin\mathcal{E},\;i\neq j,
\end{cases}
\]
\[
K_{ii}=-\sum_{j\neq i}K_{ji}.
\]
Note that in the convention adopted here, the element $K_{ij}$ ($i\neq j$)
represents the rate of transition from state $j$ to state $i$, as a
result of which the sum of elements of each column of the matrix $K$
equals zero ($\sum_{i}K_{ij}=0$), and the generator acts on the
column vector of probabilities $\mathbf{f}$ according to the equation
$\dot{\mathbf{f}}=K\mathbf{f}$. The matrix $K$ is the generator of a
continuous-time Markov process. It is non-symmetric, which complicates
the numerical analysis of its spectrum, determined by the standard
eigenvalue equation $K\phi=\lambda\phi$. The standard trick is to
transition to a symmetric matrix $S$, which has the same eigenvalues as
the matrix $K$. Define $S$ through the similarity transformation
$S=W^{-1/2}KW^{1/2}$, where $W=\mathrm{diag}(w_{1},\dots,w_{V})$.
For $i\neq j$, the elements of matrix $S$ are:
\begin{equation}
S_{ij}=\frac{1}{\sqrt{w_{i}}}K_{ij}\sqrt{w_{j}}=\nu_{0}\exp\left(-\frac{|F_{i}-F_{j}|}{2RT}\right)\;\text{for }\{i,j\}\in\mathcal{E},\label{eq_S_offdiag}
\end{equation}
and $S_{ij}=0$ for non-adjacent vertices. The diagonal elements under
this transformation are preserved:
\begin{equation}
S_{ii}=K_{ii}=-\sum_{j\neq i}K_{ji}=-\sum_{j\neq i}\nu_{0}\exp\left(-\frac{\max\{F_{i},F_{j}\}-F_{i}}{RT}\right).\label{eq_S_diag}
\end{equation}
The matrix $S$ constructed in this way is symmetric and has a spectrum
that exactly coincides with the spectrum of $K$. However, $S$ is not a
generator of a Markov process in the strict sense, since the sums of
elements in its columns are not equal to zero. The stationary
distribution of the original process (eigenvalue $\lambda_{0}=0$)
corresponds to an eigenvector of matrix $S$ of the form
$\boldsymbol{\psi}_{0}=\frac{1}{\sqrt{Z}}W^{1/2}\mathbf{1}$,
where $\mathbf{1}$ is the vector of ones, and $Z=\sum^{V}_{i=1}w_{i}$
is the total partition function of the system.

The connection between the spectra of matrices $K$ and $S$ is
established directly through the similarity transformation.
Substituting $\phi=W^{1/2}\psi$ into the original equation
$K\phi=\lambda\phi$ and left-multiplying by $W^{-1/2}$ leads to the
standard symmetric eigenvalue problem:
\begin{equation}
S\psi=\lambda\psi.\label{eq_standard_eigenvalue}
\end{equation}
Consequently, all eigenvalues $\lambda_{k}$, and hence the relaxation
times $\tau_{k}=-1/\lambda_{k}$, are preserved when passing from the
original process to the symmetrized description. The replacement of
matrix $K$ by matrix $S$ is due exclusively to the computational
efficiency of diagonalizing symmetric matrices. The eigenvectors
$\psi_{k}$ of the symmetrized matrix $S$ are orthonormal in the
standard Euclidean scalar product.

It is easy to see that the matrix $S$ is negative semidefinite. Indeed,
consider an arbitrary vector $x\in\mathbb{R}^{V}$ and expand the
quadratic form for the symmetric matrix $S$, using the definitions of
its elements (\ref{eq_S_offdiag})--(\ref{eq_S_diag}):
\[
x^{T}Sx=\sum_{i}S_{ii}x^{2}_{i}+\sum_{i\neq j}S_{ij}x_{i}x_{j}
\]
\begin{equation}
=-\sum_{i}\left(\sum_{j\neq i}K_{ji}\right)x^{2}_{i}+\sum_{\{i,j\}\in\mathcal{E}}2S_{ij}x_{i}x_{j}.\label{xSx}
\end{equation}
Using the relation between the elements of $S$ and $K$, write
$S_{ij}=\frac{1}{\sqrt{w_{i}}}K_{ij}\sqrt{w_{j}}=\frac{1}{\sqrt{w_{j}}}K_{ji}\sqrt{w_{i}}$,
from which $K_{ji}=S_{ij}\sqrt{w_{j}/w_{i}}$. Substituting this
expression into (\ref{xSx}), we obtain after elementary transformations:
\[
x^{T}Sx=-\sum_{\{i,j\}\in\mathcal{E}}S_{ij}\left(\sqrt{\frac{w_{j}}{w_{i}}}x^{2}_{i}+\sqrt{\frac{w_{i}}{w_{j}}}x^{2}_{j}-2x_{i}x_{j}\right)
\]
\begin{equation}
=-\sum_{\{i,j\}\in\mathcal{E}}S_{ij}\left(\sqrt[4]{\frac{w_{j}}{w_{i}}}x_{i}-\sqrt[4]{\frac{w_{i}}{w_{j}}}x_{j}\right)^{2}\le0.\label{eq_negative_semidef}
\end{equation}
Inequality (\ref{eq_negative_semidef}) holds by virtue of the
non-negativity of all off-diagonal elements $S_{ij}$ ($i\neq j$).
Equality to zero is achieved if and only if
$x_{i}/\sqrt{w_{i}}=x_{j}/\sqrt{w_{j}}$ for all pairs
$\{i,j\}\in\mathcal{E}$, i.e., when the vector $x$ is proportional to
$\sqrt{w}$ on each connected component of the graph $G$.

From negative semidefiniteness it follows that $S$ has a complete set
of eigenvalues $\lambda_{0},\lambda_{1},\dots,\lambda_{V-1}\in\mathbb{R}$,
satisfying the condition
$0=\lambda_{0}\ge\lambda_{1}\ge\lambda_{2}\ge\dots\ge\lambda_{V-1}$.

\section{Basins of Attraction and Kinetic Metric}

\label{sec_bas}

Let a parameter $\varepsilon_{\mathrm{eq}}>0$, called the energy
equivalence threshold, be given. We call an attraction point a
microstate $i$ for which all neighbors (in the sense of graph $G$) have
free energy not less than $F_{i}-\varepsilon_{\mathrm{eq}}$:
\[
\forall j\in\mathcal{N}(i):F_{j}\ge F_{i}-\varepsilon_{\mathrm{eq}},
\]
where $\mathcal{N}(i)$ is the set of neighbors of vertex $i$ in graph
$G$. Attraction points connected by a path in graph $G$, all of whose
edges connect microstates with a free energy difference not exceeding
$\varepsilon_{\mathrm{eq}}$, are merged into one attraction point (this
procedure allows merging degenerate minima into shallow plateaus). Each
attraction point (possibly after merging) becomes the center of a basin
of attraction.

The basin of attraction $\mathcal{B}_{a}$ corresponding to a given
attraction point $i_{a}$ is the set of all microstates $i$ for which
the gradient descent trajectory ends in $i_{a}$. The gradient descent
procedure is defined as follows. For each microstate $i$ not yet
assigned to any basin, among all its neighbors $j$ having free energy
strictly less than $F_{i}-\varepsilon_{\mathrm{eq}}$, the neighbor with
the minimum energy is chosen, and microstate $i$ is assigned to the
same basin as this neighbor. If a microstate has no neighbors with
energy strictly less than $F_{i}-\varepsilon_{\mathrm{eq}}$, it is
itself an attraction point and forms a new basin. The procedure is
repeated for all microstates in ascending order of free energy. As a
result, a family of basins of attraction
$\{\mathcal{B}_{1},\dots,\mathcal{B}_{K}\}$ is formed, where the number
$K$ is determined by the graph topology and the value of
$\varepsilon_{\mathrm{eq}}$. Each basin $\mathcal{B}_{a}$ represents a
subset of microstates from which gradient descent trajectories lead to
the same attraction point.

An important aspect of constructing basins of attraction is the problem
of connectivity of the microstate graph. Due to the stochastic nature,
the graph $G$ may decompose into several connected components. To
correctly account for this factor, we will use an analysis by connected
components, completely analogous to that implemented in
\cite{zubarev2026_rna}. Let the graph $G$ decompose into $C$ connected
components $G_{1},\dots,G_{C}$. For each component $G_{c}$ containing
at least $K_{\min}$ basins of attraction (where $K_{\min}$ is a given
threshold), a symmetrized transition rate matrix $S_{c}$ is
independently constructed (as a submatrix of the original matrix on the
indices of microstates belonging to $G_{c}$). To assess the degree of
isolation of components, a fragmentation index is introduced, defined
as
\[
f_{\mathrm{inter}}=1-\frac{\sum_{c}\binom{K_{c}}{3}}{\binom{K_{\mathrm{total}}}{3}},
\]
where $K_{c}$ is the number of basins in component $c$,
$K_{\mathrm{total}}=\sum_{c}K_{c}$ is the total number of basins in
significant components, and $\binom{K}{3}$ is the number of unordered
triples. The index $f_{\mathrm{inter}}$ takes values from 0 (all basins
belong to one component) to 1 (each basin is isolated). A high value of
$f_{\mathrm{inter}}$ indicates that significant basins of attraction
are distributed over several mutually unconnected components of the
graph. In this case, the observed ultrametric structure characterizes
the mutual arrangement of basins inside each individual component, but
does not reflect the global organization of the entire energy landscape,
since transitions between different connected components are forbidden.

Let us proceed to the construction of the kinetic metric. For each basin
of attraction $\mathcal{B}_{a}$, its vector of conditional equilibrium
distribution $\mathbf{p}_{a}\in\mathbb{R}^{V}$ is specified with
components

\begin{equation}
(\mathbf{p}_{a})_{i}=p_{a,i}=\begin{cases}
w_{i}/W_{a}, & i\in\mathcal{B}_{a},\\
0, & i\notin\mathcal{B}_{a},
\end{cases}\label{eq_p_a_i}
\end{equation}
where $W_{a}=\sum_{j\in\mathcal{B}_{a}}w_{j}$ is the partition function
of the basin.

Next, a fundamental question arises: how to correctly construct a
kinetic metric using the eigenvectors of the symmetric matrix $S$, if
the original physical process is described by the non-symmetric matrix
$K$? Direct projection of $\mathbf{p}_{a}$ onto the eigenvectors
$\boldsymbol{\psi}_{k}$ of matrix $S$ would be erroneous, since
$\boldsymbol{\psi}_{k}$ are not eigenvectors of $K$. Below we provide a
justification of exactly how the projections should be computed to
obtain a metric corresponding to the original process.

The original matrix $K$ satisfies the detailed balance condition:
$K_{ij}w_{j}=K_{ji}w_{i}$, where $w_{i}=\exp(-F_{i}/RT)$. Let us
introduce the diagonal matrix of stationary weights
$W=\mathrm{diag}(w_{1},\dots,w_{V})$. The detailed balance condition is
equivalent to the symmetry of the matrix $KW$: $(KW)^{T}=KW$, where
$W=\mathrm{diag}(w_{1},\dots,w_{V})$ is the diagonal matrix of
stationary weights. This means that the matrix $K$ is similar to the
symmetric matrix $S=W^{-1/2}KW^{1/2}$. Consider for matrix $K$ the right
eigenvectors $\boldsymbol{\phi}_{k}$, satisfying
$K\boldsymbol{\phi}_{k}=\lambda_{k}\boldsymbol{\phi}_{k}$, and the
left eigenvectors $\tilde{\boldsymbol{\phi}}_{k}$, satisfying
$\tilde{\boldsymbol{\phi}}^{T}_{k}K=\lambda_{k}\tilde{\boldsymbol{\phi}}^{T}_{k}$.
From the detailed balance condition it follows that the left and right
eigenvectors are related by a simple algebraic relation:
\[
\tilde{\boldsymbol{\phi}}_{k}=W^{-1}\boldsymbol{\phi}_{k}.
\]
Indeed, transposing the equality
$K\boldsymbol{\phi}_{k}=\lambda_{k}\boldsymbol{\phi}_{k}$, we obtain
$\boldsymbol{\phi}^{T}_{k}K^{T}=\lambda_{k}\boldsymbol{\phi}^{T}_{k}$.
Multiplying by $W^{-1}$ and using $K^{T}W^{-1}=W^{-1}K$ (which is
equivalent to detailed balance), we arrive at
$\boldsymbol{\phi}^{T}_{k}W^{-1}K=\lambda_{k}\boldsymbol{\phi}^{T}_{k}W^{-1}$,
i.e., $\tilde{\boldsymbol{\phi}}^{T}_{k}K=\lambda_{k}\tilde{\boldsymbol{\phi}}^{T}_{k}$
with $\tilde{\boldsymbol{\phi}}_{k}=W^{-1}\boldsymbol{\phi}_{k}$. The
right and left eigenvectors form a biorthogonal system:
\[
\langle\tilde{\boldsymbol{\phi}}_{k},\boldsymbol{\phi}_{\ell}\rangle=\delta_{k\ell}.
\]
An arbitrary probability distribution $\mathbf{f}(t)$ is expanded over
the right eigenvectors with coefficients computed via the scalar product
with the left eigenvectors:
\[
\mathbf{f}(t)=\sum_{k}c_{k}(t)\boldsymbol{\phi}_{k},\;c_{k}(t)=\langle\tilde{\boldsymbol{\phi}}_{k},\mathbf{f}(t)\rangle.
\]
This is the standard biorthogonal expansion. The physical projection of
macrostate $\mathbf{p}_{a}$ onto the $k$-th relaxation mode is computed
via the scalar product with the left eigenvector:
\begin{equation}
\pi^{(\mathrm{phys})}_{ak}=\langle\tilde{\boldsymbol{\phi}}_{k},\mathbf{p}_{a}\rangle=\sum_{i\in\mathcal{B}_{a}}w^{-1}_{i}\phi_{k,i}\frac{w_{i}}{W_{a}}=\frac{1}{W_{a}}\sum_{i\in\mathcal{B}_{a}}\phi_{k,i}.\label{eq_proj_def_final}
\end{equation}
where $\phi_{k,i}=\left(\boldsymbol{\phi}_{k}\right)_{i}$. Exactly
these coefficients are the physical projections and must be used in the
formula for the kinetic metric.

Computing the eigenvectors of the non-symmetric matrix $K$ for large
sparse systems is numerically unstable and computationally expensive.
Therefore, we use the symmetric matrix $S=W^{-1/2}KW^{1/2}$, whose
eigenvectors $\boldsymbol{\psi}_{k}$ are related to the right
eigenvectors of $K$ by the transformation
\[
\boldsymbol{\psi}_{k}=W^{-1/2}\boldsymbol{\phi}_{k}\;\Longleftrightarrow\;\boldsymbol{\phi}_{k}=W^{1/2}\boldsymbol{\psi}_{k}.
\]
Substituting this relation into formula (\ref{eq_proj_def_final}), we
obtain
$\pi^{(\mathrm{phys})}_{ak}=\frac{1}{W_{a}}\sum_{i\in\mathcal{B}_{a}}w^{1/2}_{i}\psi_{k,i}.$
Therefore, the physical projection is expressed through the standard
Euclidean scalar product in the symmetric basis as
\begin{equation}
\pi^{(\mathrm{phys})}_{ak}=\langle\mathbf{u}_{a},\boldsymbol{\psi}_{k}\rangle,\label{eq_final_projection}
\end{equation}
where $\mathbf{u}_{a}$ is a vector with components
\begin{equation}
(\mathbf{u}_{a})_{i}=u_{a,i}=\begin{cases}
w^{1/2}_{i}/W_{a}, & i\in\mathcal{B}_{a},\\
0, & i\notin\mathcal{B}_{a}.
\end{cases}\label{eq_u_a_i}
\end{equation}
Thus, using the vectors $\mathbf{u}_{a}$ together with the eigenvectors
$\boldsymbol{\psi}_{k}$ of matrix $S$ gives exactly the same numerical
values of projections as using the left/right pair of eigenvectors of
the original matrix $K$.

We next define the square of the kinetic distance between basins
$\mathcal{B}_{a}$ and $\mathcal{B}_{b}$ as the weighted sum of squared
differences of physical projections:
\begin{equation}
D^{2}(\mathcal{B}_{a},\mathcal{B}_{b})=\sum^{n_{\mathrm{phys}}}_{k=1}\frac{1}{|\lambda_{k}|}\,\left(\pi^{(\mathrm{phys})}_{ak}-\pi^{(\mathrm{phys})}_{bk}\right)^{2}.\label{eq_mahalanobis_macro}
\end{equation}
The form (\ref{eq_mahalanobis_macro}) is nothing other than the square
of the Mahalanobis distance \cite{Mahalanobis}. Since the eigenvalues
$\lambda_{k}$ of matrices $K$ and $S$ coincide (they are invariant
under similarity transformation), and the projections
$\pi^{(\mathrm{phys})}_{ak}$ are computed correctly via
(\ref{eq_final_projection}), this distance (\ref{eq_mahalanobis_macro})
is a kinetic characteristic of the original Markov process. We note that
in this case the replacement of the non-symmetric problem by a symmetric
one is of a purely computational nature and does not change the physical
content of the metric.

The weights $1/|\lambda_{k}|$ in formula (\ref{eq_mahalanobis_macro})
have the following physical interpretation. The quantity $|\lambda_{k}|$
characterizes the rate of the $k$-th relaxation process -- the slower
the process (the smaller $|\lambda_{k}|$), the greater the contribution
of the difference of projections onto the corresponding mode to the
distance between macrostates. This reflects the fact that transitions
between macrostates having different projections onto slow modes are
hindered by high energy barriers. The coefficients
$\pi^{(\mathrm{phys})}_{ak}$ are the coordinates of the vector
$\mathbf{u}_{a}$ in the basis of eigenvectors
$\{\boldsymbol{\psi}_{k}\}^{V-1}_{k=1}$ of matrix $S$. Physically,
$\pi^{(\mathrm{phys})}_{ak}$ determines the amplitude with which the
$k$-th relaxation mode is represented in the probability distribution
concentrated on basin $\mathcal{B}_{a}$: the larger
$|\pi^{(\mathrm{phys})}_{ak}|$, the greater the contribution of the
$k$-th mode to the dynamics of the transition from this basin. The
quantity $|\lambda_{k}|^{-1}$ is the relaxation time of the $k$-th
mode. The distance $D(\mathcal{B}_{a},\mathcal{B}_{b})$ is computed
through the differences of projections
$\pi^{(\mathrm{phys})}_{ak}-\pi^{(\mathrm{phys})}_{bk}$, since it is
precisely the difference in the amplitudes of modes between the two
basins that determines the kinetic non-equivalence of their relaxation
properties: the stronger the difference in projections onto the slow
modes, the greater the effective transition time between the basins.

The sum of squared projections
$\sum^{V-1}_{k=1}(\pi^{(\mathrm{phys})}_{ak})^{2}$ is not equal to
unity and depends on the basin, since the vectors $\mathbf{u}_{a}$ are
not normalized: $\|\mathbf{u}_{a}\|^{2}=1/W_{a}$. From the
orthonormality of the basis $\{\boldsymbol{\psi}_{k}\}^{V-1}_{k=0}$ and
the fact that the normalized zero eigenvector is
$\boldsymbol{\psi}_{0}=W^{1/2}\mathbf{1}/\sqrt{Z}$, the projection
onto the zero mode equals
$\langle\mathbf{u}_{a},\boldsymbol{\psi}_{0}\rangle=1/\sqrt{Z}$.
According to Parseval's identity:
\[
\|\mathbf{u}_{a}\|^{2}=\langle\mathbf{u}_{a},\boldsymbol{\psi}_{0}\rangle^{2}+\sum^{V-1}_{k=1}(\pi^{(\mathrm{phys})}_{ak})^{2}\;\implies\;\frac{1}{W_{a}}=\frac{1}{Z}+\sum^{V-1}_{k=1}(\pi^{(\mathrm{phys})}_{ak})^{2},
\]
from which it follows that
\[
\sum^{V-1}_{k=1}(\pi^{(\mathrm{phys})}_{ak})^{2}=\frac{1}{W_{a}}-\frac{1}{Z}.
\]
Different values of this sum for different basins reflect their unequal
kinetic prominence. The smaller the statistical weight $W_{a}$ (i.e.,
the narrower and deeper the basin), the larger
$\sum_{k}(\pi^{(\mathrm{phys})}_{ak})^{2}$, which means the dominance
of slow modes in the dynamics of the transition from such a basin.
Conversely, basins with large $W_{a}$ (wide and shallow) have a small
sum of squared projections, since their dynamics are concentrated in
fast modes. The complete invariant, independent of the basin, is
\begin{equation}
W_{a}\left(\langle\mathbf{u}_{a},\boldsymbol{\psi}_{0}\rangle^{2}+\sum^{V-1}_{k=1}(\pi^{(\mathrm{phys})}_{ak})^{2}\right)=W_{a}\left(\frac{1}{Z}+\frac{1}{W_{a}}-\frac{1}{Z}\right)=1.\label{eq_inv}
\end{equation}
Relation (\ref{eq_inv}) means that upon multiplication by the
statistical weight, the sum of all squared projections (including the
zero mode) is normalized to unity for any basin.

We introduce on the set of found basins of attraction
$\left\{ \mathcal{B}_{a}\right\} $ an equivalence relation $\sim_{m}$,
defined by the equality of projections onto the subspace $\mathcal{V}$:
\[
\mathcal{B}_{a}\sim_{m}\mathcal{B}_{b}\;\Longleftrightarrow\;\pi^{(\mathrm{phys})}_{a}=\pi^{(\mathrm{phys})}_{b}\;\Longleftrightarrow\;D(\mathcal{B}_{a},\mathcal{B}_{b})=0.
\]
Denote by $\left\{ \mathcal{B}_{a}\right\} /\sim_{m}$ the quotient
set of equivalence classes under this relation. Then one can prove (see
\cite{zubarev2026_rna} for details) that the function
(\ref{eq_mahalanobis_macro}) is a metric on the quotient set
$\left\{ \mathcal{B}_{a}\right\} /\!\sim_{m}$. We emphasize that this
metric is not automatically an ultrametric, since the Euclidean metric
in the weighted space of projections does not necessarily satisfy the
strong triangle inequality. Therefore, checking the ultrametricity of
(\ref{eq_mahalanobis_macro}) is a substantive task, the result of which
is determined by the real structure of the energy landscape.

\section{Ultrametricity Criterion}

\label{sec_ultra}

Let $\{\mathcal{B}_{1},\dots,\mathcal{B}_{K_{c}}\}$ be the set of
basins in the connected component $c$. To test the ultrametricity of
this component, we will analyze all unordered triples of distinct
elements from $\{\mathcal{B}_{1},\dots,\mathcal{B}_{K_{c}}\}$.

Let $(\mathcal{B}^{\prime},\mathcal{B}^{\prime\prime},\mathcal{B}^{\prime\prime\prime})$
be a triple of distinct elements from
$\{\mathcal{B}_{1},\dots,\mathcal{B}_{K_{c}}\}$. Order the three
pairwise distances in ascending order:
$d_{\min}\le d_{\text{med}}\le d_{\max}$. Fix two real parameters
$\varepsilon$ and $\delta$, satisfying the condition
$0<\varepsilon<\delta$.

Triples of elements are classified according to the following rules. A
triple $(\mathcal{B}^{\prime},\mathcal{B}^{\prime\prime},\mathcal{B}^{\prime\prime\prime})$
is called trivially ultrametric if the condition
\begin{equation}
\frac{d_{\max}-d_{\min}}{d_{\min}}\le\varepsilon.\label{eq_trivial}
\end{equation}
is satisfied. In the case $d_{\min}=0$, the triple is considered
trivially ultrametric only if $d_{\max}=0$. Trivial ultrametricity
means that all three distances are approximately equal. Such a
situation can arise, for example, with a random choice of points in a
high-dimensional space \cite{zubarev2014,zubarev2017} and does not
necessarily indicate the presence of a hierarchical structure. A triple
$(\mathcal{B}^{\prime},\mathcal{B}^{\prime\prime},\mathcal{B}^{\prime\prime\prime})$
is called nontrivially ultrametric if two conditions are satisfied
simultaneously:
\begin{equation}
\frac{d_{\max}-d_{\text{med}}}{d_{\text{med}}}\le\varepsilon,\;\frac{d_{\text{med}}-d_{\min}}{d_{\text{med}}}>\delta.\label{eq_nontrivial}
\end{equation}
Nontrivial ultrametricity corresponds to the case when the two largest
distances are approximately equal, and the third is substantially
smaller. It is precisely this structure of triples that is typical for
ultrametric spaces isomorphic to the set of leaves of a multilevel
rooted tree. A triple that satisfies neither condition (\ref{eq_trivial})
nor condition (\ref{eq_nontrivial}) is called non-ultrametric.

Let $\mathcal{T}$ be the set of all unordered triples of distinct
elements from $\{\mathcal{B}_{1},\dots,\mathcal{B}_{K_{c}}\}$. Its
cardinality is equal to $\binom{K_{c}}{3}$. Let
$\mathcal{T}_{\text{nt}}\subset\mathcal{T}$ be the subset of
nontrivially ultrametric triples. The degree of nontrivial
ultrametricity of the space of component
$\{\mathcal{B}_{1},\dots,\mathcal{B}_{K_{c}}\}$ is defined as
\begin{equation}
u_{\text{nt}}=\frac{|\mathcal{T}_{\text{nt}}|}{|\mathcal{T}|}\times100\%.\label{eq_u_nt}
\end{equation}
The degree of trivial ultrametricity $u_{\text{tr}}$ and the degree of
non-ultrametricity $u_{\text{non}}$ are defined similarly, with
$u_{\text{nt}}+u_{\text{tr}}+u_{\text{non}}=100\%$. The degrees of
trivial ultrametricity, nontrivial ultrametricity, and the degree of
non-ultrametricity of the entire set of basins are defined as weighted
sums of the corresponding degrees of ultrametricity of its connected
components.

Note that in continuous metrics, exact equalities required by
conditions (\ref{eq_trivial}) and (\ref{eq_nontrivial}) at $\delta=0$,
$\varepsilon=0$ have zero probability. Therefore, in computational
experiments, approximate inequalities with specific values of
$\varepsilon$ and $\delta$ are used, chosen based on the required
rigor of classification.

\section{Computational Results}

\label{sec_results}

To perform the calculations, we developed a specialized program in the
Python interpreter, the source code of which is in open access
\cite{zubarev2026_github}. Using this program, to test the hypothesis
of the ultrametric organization of random energy landscapes, a series
of computational experiments was conducted with the following fixed
parameters: number of vertices $V=5000$, number of edges $E=5000$
(which corresponds to an average degree $\langle k\rangle=2.0$ and a
theoretical fraction of the giant component $S\approx0.7968$); Kramers
temperature $T=300$ K ($RT\approx2.49$ kJ/mol); energy equivalence
threshold $\varepsilon_{\mathrm{eq}}=3.0$ kJ/mol; minimum basin size
$s_{\min}=10$ vertices; minimum number of basins in a connected
component $K_{\min}=5$; ultrametric triple classification parameters
$\varepsilon=0.05$, $\delta=0.10$; number of requested eigenmodes
$n_{\mathrm{modes}}=50$; spectral gap threshold $\theta=10^{6}$; number
of independent realizations for each parameter set $N=10$. The varied
parameter was the width of the interval of the uniform distribution of
free energies $\Delta F=F_{\max}-F_{\min}$ (with fixed $F_{\min}=0$).
Seven values of the free energy distribution interval were investigated:
$\Delta F=10$, $20$, $50$, $100$, $200$, $500$, and $1000$ kJ/mol.

Analysis of the connected components for all realizations showed results
fully consistent with the theory of Erdős--Rényi random graphs. The
average size of the giant component was $4000.5\pm34.8$ vertices with
an average number of edges inside it of $4801.3\pm31.9$. The fraction
of vertices in the giant component $S_{\mathrm{obs}}\approx0.800$
practically coincides with the theoretical value $S=0.7968$. The
remaining vertices decomposed into many small components, the size of
which did not exceed a few tens of vertices, and which were
automatically discarded by the algorithm as not containing a sufficient
number of basins. The fragmentation index $f_{\mathrm{inter}}$ in all
successful realizations was equal to zero, which indicates that all
significant basins belonged to a single giant connected component.

The main results, averaged over ten independent realizations for each
value of $\Delta F$, are presented in Table \ref{tab_main}.

\begin{table}[H]
\centering \caption{Dependence of the degree of nontrivial ultrametricity $u_{\mathrm{nt}}$
and the average number of basins $\langle K\rangle$ on the width of the
energy distribution $\Delta F$. Presented are mean values $\pm$ standard
deviation over ten realizations.}
\label{tab_main}
\global\long\def\arraystretch{1.2}%
\begin{tabular}{|c|c|c|c|}
\hline
$\Delta F$, kJ/mol  & $\langle K\rangle$  & $u_{\mathrm{nt}}$, \%  & $u_{\mathrm{non}}$, \% \tabularnewline
\hline
$10$  & $54.9\pm7.0$  & $42.06\pm9.88$  & $57.02\pm9.64$ \tabularnewline
$20$  & $31.9\pm3.6$  & $60.65\pm11.48$  & $39.25\pm11.49$ \tabularnewline
$50$  & $24.5\pm3.8$  & $81.03\pm6.82$  & $16.72\pm8.70$ \tabularnewline
$100$  & $26.1\pm2.2$  & $82.60\pm6.32$  & $16.85\pm6.18$ \tabularnewline
$200$  & $28.5\pm2.3$  & $86.53\pm7.63$  & $13.43\pm7.65$ \tabularnewline
$500$  & $29.5\pm2.9$  & $90.63\pm4.95$  & $9.05\pm5.19$ \tabularnewline
$1000$  & $30.3\pm3.0$  & $95.91\pm1.69$  & $3.65\pm1.88$ \tabularnewline
\hline
\end{tabular}
\end{table}

As can be seen from Table \ref{tab_main}, a clear monotonic dependence
of the degree of nontrivial ultrametricity on the width of the energy
distribution is observed. For small values of $\Delta F=10$ and
$20$ kJ/mol, the kinetic metric reveals moderate ultrametricity: about
$42\%$ and $61\%$ of triples of basins are classified as nontrivially
ultrametric. At $\Delta F=50$ kJ/mol (which corresponds to $\sim20RT$),
the kinetic metric reveals a high degree of ultrametricity: about $81\%$
of triples are classified as nontrivially ultrametric. With an increase
of $\Delta F$ to $100$ kJ/mol, the fraction of nontrivially ultrametric
triples reaches a plateau at a level of $\approx83\%$. At
$\Delta F=200$ kJ/mol, another increase occurs: $u_{\mathrm{nt}}$
rises to $86.5\%$, and the fraction of non-ultrametric triples falls to
$13\%$. A further increase of $\Delta F$ to $500$ kJ/mol and
$1000$ kJ/mol is accompanied by a continuation of the monotonic growth
of $u_{\mathrm{nt}}$ to $90.6\%$ and $95.9\%$, respectively. The
fraction of non-ultrametric triples at $\Delta F=1000$ kJ/mol drops to
$3.6\%$, which indicates that the system is entering the asymptotic
regime.

The average number of basins of attraction $\langle K\rangle$ also
shows a tendency to grow with increasing $\Delta F$ (from $24.5$ to
$30.3$); however, this growth is relatively weak compared to the change
in $u_{\mathrm{nt}}$. This observation indicates that the increase in
the degree of ultrametricity is not a simple consequence of the
fragmentation of the landscape into a larger number of basins, but
reflects a fundamental change in the structure of the kinetic metric.

To illustrate the nature of nontrivially ultrametric triples, Table
\ref{tab_triangles} presents characteristic examples of triples of
distances for $\Delta F=50$, $\Delta F=200$, and $\Delta F=1000$ kJ/mol
(taken from realizations with maximum $u_{\mathrm{nt}}$).

\begin{table}[H]
\centering \caption{Examples of nontrivially ultrametric triples of distances for three values
of $\Delta F$. In each cell, triples $(d_{\max},d_{\mathrm{med}},d_{\min})$
are presented from top to bottom.}
\label{tab_triangles}
\global\long\def\arraystretch{1.2}%
\begin{tabular}{|c|c|c|}
\hline
$\Delta F=50$ kJ/mol  & $\Delta F=200$ kJ/mol  & $\Delta F=1000$ kJ/mol \tabularnewline
\hline
$6.17\cdot10^{18}$  & $2.03\cdot10^{21}$  & $2.55\cdot10^{63}$ \tabularnewline
$6.17\cdot10^{18}$  & $2.03\cdot10^{21}$  & $2.55\cdot10^{63}$ \tabularnewline
$2.11\cdot10^{11}$  & $1.10\cdot10^{13}$  & $8.47\cdot10^{53}$ \tabularnewline
\hline
$2.11\cdot10^{11}$  & $1.10\cdot10^{13}$  & $2.55\cdot10^{63}$ \tabularnewline
$2.11\cdot10^{11}$  & $1.10\cdot10^{13}$  & $2.55\cdot10^{63}$ \tabularnewline
$2.58\cdot10^{9}$  & $1.98\cdot10^{10}$  & $4.68\cdot10^{45}$ \tabularnewline
\hline
$2.12\cdot10^{11}$  & $1.17\cdot10^{16}$  & $2.06\cdot10^{70}$ \tabularnewline
$2.11\cdot10^{11}$  & $1.17\cdot10^{16}$  & $2.06\cdot10^{70}$ \tabularnewline
$1.95\cdot10^{10}$  & $1.10\cdot10^{13}$  & $2.55\cdot10^{63}$ \tabularnewline
\hline
$2.18\cdot10^{11}$  & $4.11\cdot10^{16}$  & $2.55\cdot10^{63}$ \tabularnewline
$2.11\cdot10^{11}$  & $4.11\cdot10^{16}$  & $2.55\cdot10^{63}$ \tabularnewline
$5.60\cdot10^{10}$  & $1.10\cdot10^{13}$  & $1.41\cdot10^{32}$ \tabularnewline
\hline
$1.15\cdot10^{12}$  & $3.45\cdot10^{17}$  & $5.19\cdot10^{79}$ \tabularnewline
$1.13\cdot10^{12}$  & $3.45\cdot10^{17}$  & $5.19\cdot10^{79}$ \tabularnewline
$2.11\cdot10^{11}$  & $1.10\cdot10^{13}$  & $2.55\cdot10^{63}$ \tabularnewline
\hline
\end{tabular}
\end{table}

A qualitative difference in the structure of distances between regimes
with small and large $\Delta F$ is noteworthy. At $\Delta F=50$ kJ/mol,
the absolute values of distances are on the order of $10^{11}$--$10^{18}$,
and the ratios $d_{\max}/d_{\min}$ reach several orders of magnitude. At
$\Delta F=200$ kJ/mol, the distances increase to the order of
$10^{13}$--$10^{21}$, with the two largest distances $d_{\max}$ and
$d_{\mathrm{med}}$ coinciding with high accuracy, while $d_{\min}$ is
two to three orders of magnitude smaller. At $\Delta F=1000$ kJ/mol,
the distances reach astronomical values on the order of
$10^{63}$--$10^{79}$; here $d_{\max}$ and $d_{\mathrm{med}}$ are
practically indistinguishable, and $d_{\min}$ is tens of orders of
magnitude smaller. Such a structure is a direct indication that at large
$\Delta F$, the kinetic metric approximates ever more accurately the
strict ultrametric, in which for any nontrivial triple the two largest
distances are exactly equal.

The reader may be surprised that the absolute values of the distances
$D$ reach extremely large magnitudes (on the order of $10^{18}$ and even
$10^{63}$). It is important to emphasize that this is not a numerical
artifact or computational error, but represents a fundamental physical
consequence of the Arrhenius law and Kramers theory. The kinetic
distance $D$ is inversely proportional to the square root of the
absolute value of the eigenvalue of the rate matrix
($D\sim1/\sqrt{|\lambda|}$), which physically characterizes the
relaxation time (or mean committor time) between macrostates. According
to Kramers theory, the rate of transition through a minimax barrier of
height $U^{*}$ is exponentially suppressed:
$|\lambda|\sim\exp(-U^{*}/RT)$. Consequently, the kinetic distance grows
exponentially with the barrier height and also accounts for entropic
factors (statistical weights of basins):
\[
D\sim\exp\left(\frac{U^{*}}{2RT}\right).
\]
In our computational experiments, with a distribution width
$\Delta F=50$ kJ/mol, the distances between basins already reach
magnitudes of order $10^{18}$. When the interval increases to
$\Delta F=1000$ kJ/mol at temperature $T=300$ K
($RT\approx2.49$ kJ/mol), typical heights of separating barriers
$U^{*}$ amount to hundreds of kJ/mol. Thus, a barrier of height
$U^{*}\approx730$ kJ/mol gives an asymptotic estimate
$D\sim\exp(730/4.98)\sim10^{63}$, which exactly corresponds to the
observed values. The waiting times for such transitions at room
temperature exceed the age of the Universe by many orders of magnitude,
which means absolute kinetic isolation of the basins. Thus, the gigantic
values of $D$ are a direct numerical confirmation of the Limit Theorem
formulated in Section \ref{sec_limit_theorem}: the logarithmic
asymptotics of the metric $\frac{1}{\beta}\ln D$ exactly corresponds to
the minimax barrier height $\frac{1}{2}\rho^{(0)}$. For practical
analysis and visualization of such landscapes (for example, when
constructing dendrograms), the standard and mathematically correct
procedure is to use the logarithmic scale of distances ($\ln D$), which
is fully consistent with the traditions of spin glass physics and
$p$-adic analysis.

\section{Limit Theorem on the Convergence of the Kinetic Metric to the Single-Linkage Metric}

\label{sec_limit_theorem}

To analyze the asymptotic behavior of the kinetic metric, it is
convenient to pass to dimensionless variables. In the physical model
(Section \ref{sec_kramers}), the transition rates are determined by the
formula
$k_{i\to j}\propto\exp\!\bigl(-\frac{\max(F_{i},F_{j})-F_{i}}{RT}\bigr)$,
where $F_{i}$ are the free energies in dimensional units (kJ/mol). Let
us introduce a characteristic energy scale $F_{0}$ (for example, the
width of the distribution interval $\Delta F$) and define dimensionless
energies $\tilde{F}_{i}=F_{i}/F_{0}$. Then the exponent takes the form
$-\frac{F_{0}}{RT}(\max(\tilde{F}_{i},\tilde{F}_{j})-\tilde{F}_{i})$.
Denoting the dimensionless parameter $\beta=F_{0}/RT$, we obtain a
mathematical formula in which fixed dimensionless energies
$\tilde{F}_{i}$ are multiplied by the parameter $\beta$. The limit
transition $\beta\to\infty$ for fixed $\tilde{F}_{i}$ physically
means making the scale of energy fluctuations $F_{0}\sim\Delta F$ tend
to infinity at constant temperature $T$, but not cooling the system to
absolute zero. Below, for brevity, we omit the tildes over the
dimensionless energies $F_{i}$ and use the parameter $\beta$ precisely
in this scaling sense.

Let us formulate and prove the main theoretical result of the work. Fix
a connected graph $G=(\mathcal{M},\mathcal{E})$ with a set of vertices
$\mathcal{M}=\{1,\dots,V\}$. To each vertex $i$, assign a non-negative
number $F_{i}\ge0$, which we will call the free energy. Without loss of
generality, we consider all quantities $F_{i}$ to be pairwise distinct.
Introduce the dimensionless scaling parameter $\beta>0$ and consider a
family of Kramers generators $K(\beta)$, which we define by specifying
its off-diagonal elements
\[
K_{ij}(\beta)=\exp\!\bigl(-\beta(\max(F_{i},F_{j})-F_{j})\bigr),\;\{i,j\}\in\mathcal{E},
\]
and diagonal elements
$K_{ii}(\beta)=-\sum_{j\neq i}K_{ji}(\beta)$ (here the
pre-exponential factor $\nu_{0}$ is omitted, as it does not affect the
logarithmic asymptotics). The matrix $K(\beta)$ satisfies the detailed
balance condition $K_{ij}w_{j}=K_{ji}w_{i}$ with stationary Boltzmann
weights $w_{i}(\beta)=\exp(-\beta F_{i})$.

For the correct spectral decomposition of the non-symmetric matrix $K$,
we use the biorthogonal system of left and right eigenvectors. The right
eigenvectors $\boldsymbol{\phi}_{k}$ satisfy the equation
$K\boldsymbol{\phi}_{k}=\lambda_{k}\boldsymbol{\phi}_{k}$, and the
left eigenvectors $\tilde{\boldsymbol{\phi}}_{k}$ satisfy the equation
$\tilde{\boldsymbol{\phi}}^{T}_{k}K=\lambda_{k}\tilde{\boldsymbol{\phi}}^{T}_{k}$.
From the detailed balance condition it follows that the left and right
eigenvectors are related by
$\tilde{\boldsymbol{\phi}}_{k}=W^{-1}\boldsymbol{\phi}_{k}$, where
$W=\mathrm{diag}(w_{1},\dots,w_{V})$. They form a biorthogonal basis:
$\langle\tilde{\boldsymbol{\phi}}_{k},\boldsymbol{\phi}_{m}\rangle=\delta_{km}$.

The partition of the set of vertices into basins of attraction does not
depend on $\beta$. Fix this partition
$\{\mathcal{B}_{1},\dots,\mathcal{B}_{K}\}$. For each basin
$\mathcal{B}_{a}$, define the macrostate through the conditional
equilibrium probability distribution $\mathbf{p}_{a}$ with components
(\ref{eq_p_a_i}). The physical projection of macrostate $\mathbf{p}_{a}$
onto the $k$-th relaxation mode
$\langle\tilde{\boldsymbol{\phi}}_{k},\mathbf{p}_{a}\rangle=\pi^{(\mathrm{phys})}_{ak}\equiv c_{ak}$
is (\ref{eq_proj_def_final}). Write the kinetic metric on the set of
basins in the form
\[
D^{2}_{\beta}(\mathcal{B}_{a},\mathcal{B}_{b})=\sum^{V-1}_{k=1}\frac{1}{|\lambda_{k}(\beta)|}\,\bigl(c_{ak}-c_{bk}\bigr)^{2}.
\]

Let us give a definition of the single-linkage metric. Without loss of
generality, we will assume that $\min_{i\in\mathcal{M}}F_{i}=0$. Indeed,
the generator matrix $K$ is invariant under a simultaneous shift of all
free energies by an arbitrary constant $C$, since the transition rates
depend only on energy differences. The square of the kinetic metric
$D^{2}_{\beta}$ under such a shift is multiplied by a global factor
$e^{\beta C}$ (and the metric itself $D_{\beta}$ is multiplied by
$e^{\beta C/2}$), which leads to an additive shift of the logarithmic
asymptotics by $C/2$. Since the single-linkage distance $\rho^{(0)}$
under a shift of energies by $C$ also increases by $C$, both sides of
the theorem formulated below are shifted by the same amount $C/2$.
Hence, the choice of normalization $\min F_{i}=0$ does not limit the
generality of the result. Next, for each edge $\{i,j\}\in\mathcal{E}$,
define its weight
\[
U_{ij}=\max(F_{i},F_{j}).
\]
For an arbitrary path $\gamma=(v_{0},v_{1},\dots,v_{L})$ in graph $G$,
its height is defined as the quantity
\[
h(\gamma)=\max_{0\le\ell<L}U_{v_{\ell}v_{\ell+1}}.
\]
The single-linkage distance between vertices $p$ and $q$ is defined as
\[
\rho^{(0)}(p,q)=\min_{\gamma:p\to q}h(\gamma).
\]
On the set of basins, the distance is given by the formula
\[
\rho^{(0)}(\mathcal{B}_{a},\mathcal{B}_{b})=\min_{p\in\mathcal{B}_{a},\,q\in\mathcal{B}_{b}}\rho^{(0)}(p,q)
\]
for $a\neq b$ and $\rho^{(0)}(\mathcal{B}_{a},\mathcal{B}_{a})=0$.
The function $\rho^{(0)}$ is an ultrametric: symmetry is obvious, and
the strong triangle inequality
\[
\rho^{(0)}(\mathcal{B}_{a},\mathcal{B}_{b})\le\max\!\bigl(\rho^{(0)}(\mathcal{B}_{a},\mathcal{B}_{c}),\rho^{(0)}(\mathcal{B}_{c},\mathcal{B}_{b})\bigr)
\]
follows from the fact that the concatenation of optimal paths
$\mathcal{B}_{a}\to\mathcal{B}_{c}$ and
$\mathcal{B}_{c}\to\mathcal{B}_{b}$ (connected inside basin
$\mathcal{B}_{c}$ by a path, all of whose edges, by construction of the
basins, have weight strictly less than the height of any separating
barrier) gives a path from $\mathcal{B}_{a}$ to $\mathcal{B}_{b}$ whose
height does not exceed the maximum of the heights of the two original
paths.

Define also the bottom energy of basin $\mathcal{B}_{a}$ as the
minimum value of the base free energy on the microstates of this basin:
\[
F_{\min}(\mathcal{B}_{a})=\min_{i\in\mathcal{B}_{a}}F_{i}.
\]

Let $h(\gamma)=\max_{e\in\gamma}U_{e}$ be the height of an arbitrary
simple path $\gamma$ between basins $\mathcal{B}_{a}$ and
$\mathcal{B}_{b}$. A path is called optimal if $h(\gamma)=\rho$, and
suboptimal if $h(\gamma)>\rho$. Since the graph $G$ is finite, the set
of heights of all paths is finite. Define the barrier gap $\delta_{ab}$
between basins $\mathcal{B}_{a}$ and $\mathcal{B}_{b}$ as the minimum
difference between the height of any suboptimal path and the minimax
barrier:

\[
\delta_{ab}=\min_{\gamma:\,h(\gamma)>\rho}\bigl(h(\gamma)-\rho\bigr).
\]
By virtue of the finiteness of the graph and the strictness of the
inequality $h(\gamma)>\rho$, the quantity $\delta_{ab}$ is always
strictly positive.

\textbf{Theorem (Asymptotic expansion of the kinetic metric).}
Let $\rho=\rho^{(0)}(\mathcal{B}_{a},\mathcal{B}_{b})$ be the
single-linkage distance and $\delta=\delta_{ab}$ be the barrier gap
between basins $\mathcal{B}_{a}$ and $\mathcal{B}_{b}$. Let
$G_{<\rho}$ be the subgraph of graph $G$ containing all vertices of the
graph and only edges with weight strictly less than $\rho$, and
$\mathcal{E}^{*}$ be the set of edges of the original graph with weight
equal to $\rho$, connecting different connected components of subgraph
$G_{<\rho}$. Then there exists a positive constant $\mathcal{R}^{*}_{ab}$,
depending on the pair of basins $(\mathcal{B}_{a},\mathcal{B}_{b})$
and determined exclusively by the local topology of the edges from
$\mathcal{E}^{*}$ and the structure of the connected components of
subgraph $G_{<\rho}$, such that as $\beta\to\infty$ the square of the
kinetic metric admits the following exact asymptotic expansion:
\begin{equation}
D^{2}_{\beta}(\mathcal{B}_{a},\mathcal{B}_{b})=\exp(\beta\rho)\left[\mathcal{R}^{*}_{ab}+O\!\left(e^{-\beta\delta}\right)\right].\label{eq_exact_asymptotics}
\end{equation}
As a direct consequence, the logarithmic asymptotics holds:
\[
\lim_{\beta\to\infty}\frac{1}{\beta}\ln D_{\beta}(\mathcal{B}_{a},\mathcal{B}_{b})=\frac{1}{2}\rho.
\]

Note that although the physical principle of dominance of minimax
barriers is well known in the theory of large deviations for continuous
diffusion processes (Freidlin--Wentzell theory \cite{Freidlin_Wentzell}),
the rigorous derivation of the asymptotics (\ref{eq_exact_asymptotics})
directly from the spectral definition of the kinetic metric on a
discrete graph is a non-trivial mathematical problem that cannot be
reduced to classical results. The proof of the formulated theorem relies
on the apparatus of linear algebra, graph theory, and variational
calculus. We will sequentially move from the spectral definition of the
metric to a variational representation, and then obtain asymptotic
lower and upper bounds with the extraction of the leading term.

\textbf{Lemma 1 (Variational representation of the kinetic metric).}
The square of the kinetic metric
$D^{2}_{\beta}(\mathcal{B}_{a},\mathcal{B}_{b})$ admits the following
representation:
\begin{equation}
D^{2}_{\beta}(\mathcal{B}_{a},\mathcal{B}_{b})=\max_{\mathbf{h}\in\mathbb{R}^{V}:\,\langle\mathbf{w},\mathbf{h}\rangle=0}\frac{\bigl(\langle\mathbf{p}_{a},\mathbf{h}\rangle-\langle\mathbf{p}_{b},\mathbf{h}\rangle\bigr)^{2}}{\mathcal{E}(\mathbf{h},\mathbf{h})},\label{eq_Lemma_1}
\end{equation}
where $\mathbf{w}=(w_{1},\dots,w_{V})^{T}$, $w_{i}=\exp(-\beta F_{i})$,
the vectors $\mathbf{p}_{a}$ and $\mathbf{p}_{b}$ are defined by their
components $(p_{a})_{i}=w_{i}/W_{a}$ for $i\in\mathcal{B}_{a}$ (and
zeros outside $\mathcal{B}_{a}$), and $\mathcal{E}(\mathbf{h},\mathbf{h})$
is a Dirichlet quadratic form of the type
\begin{equation}
\mathcal{E}(\mathbf{h},\mathbf{h})=\sum_{\{i,j\}\in\mathcal{E}}C_{ij}(h_{i}-h_{j})^{2},\;C_{ij}=\exp\!\bigl(-\beta\max(F_{i},F_{j})\bigr).\label{eq_E_h}
\end{equation}

\textbf{Proof of Lemma 1. } Introduce the symmetric matrix $L=-S$. From
the negative semidefiniteness of $S$, proved in Section \ref{sec_kramers},
it follows that $L$ is a positive semidefinite matrix. The kernel of
matrix $L$ is one-dimensional and is spanned by the normalized
eigenvector $\boldsymbol{\psi}_{0}=\frac{1}{\sqrt{Z}}W^{1/2}\mathbf{1}$
with components $\frac{\sqrt{w_{i}}}{\sqrt{Z}}$, where
$Z=\sum^{V}_{i=1}w_{i}$ is the total partition function. In accordance
with the standard theory of spectral decomposition of symmetric
matrices, for matrix $L$ we define the Moore--Penrose pseudoinverse
matrix $L^{+}=\sum^{V-1}_{k=1}\frac{1}{\mu_{k}}\boldsymbol{\psi}_{k}\boldsymbol{\psi}^{T}_{k}$,
where $\mu_{k}=-\lambda_{k}>0$ are the non-zero eigenvalues of $L$, and
$\boldsymbol{\psi}_{k}$ ($k\ge1$) are its corresponding orthonormal
eigenvectors. It is easy to see that the kinetic metric can be
represented in the form
$D^{2}_{\beta}=(\mathbf{u}_{a}-\mathbf{u}_{b})^{T}L^{+}(\mathbf{u}_{a}-\mathbf{u}_{b})$,
where the vector $\mathbf{u}_{a}$ is defined by formula (\ref{eq_u_a_i})
or
\begin{equation}
D^{2}_{\beta}=\mathbf{y}^{T}L^{+}\mathbf{y},\;\text{where }\mathbf{y}=W^{-1/2}\mathbf{p}_{a}-W^{-1/2}\mathbf{p}_{b}.\label{eq_D_L}
\end{equation}
where the vector $\mathbf{p}_{a}$ is defined by formula (\ref{eq_p_a_i}).
The vector $\mathbf{y}$ is orthogonal to the kernel of $L$, since
$\langle\mathbf{y},\boldsymbol{\psi}_{0}\rangle=\frac{1}{\sqrt{Z}}\left(\sum_{i\in\mathcal{B}_{a}}\frac{w_{i}}{W_{a}}-\sum_{j\in\mathcal{B}_{b}}\frac{w_{j}}{W_{b}}\right)=\frac{1}{\sqrt{Z}}(1-1)=0$.
Consequently, $\mathbf{y}$ belongs to the image of matrix $L$
($\mathbf{y}\in\mathrm{Im}(L)$).

We next use a standard generalization of the Rayleigh--Ritz principle
for positive semidefinite matrices and their pseudoinverses (see, e.g.,
\cite{Horn_Johnson}). Namely, we show that for matrix $L$, a variational
principle holds, which states that for any
$\mathbf{y}\in\mathrm{Im}(L)$, $\mathbf{y}\neq\mathbf{0}$,
\begin{equation}
\mathbf{y}^{T}L^{+}\mathbf{y}=\max_{\mathbf{x}\in\mathrm{Im}(L),\,\mathbf{x}\neq\mathbf{0}}\frac{(\mathbf{y}^{T}\mathbf{x})^{2}}{\mathbf{x}^{T}L\mathbf{x}}.\label{eq_var_principle}
\end{equation}
(If $\mathbf{y}=\mathbf{0}$, then both sides are identically zero, and
the statement is trivial.)

We obtain an upper bound for (\ref{eq_var_principle}). Expand the
vectors $\mathbf{x},\mathbf{y}\in\mathrm{Im}(L)$ in the basis of
eigenvectors $\{\boldsymbol{\psi}_{k}\}^{V-1}_{k=1}$:
$\mathbf{y}=\sum^{V-1}_{k=1}d_{k}\boldsymbol{\psi}_{k}$ and
$\mathbf{x}=\sum^{V-1}_{k=1}c_{k}\boldsymbol{\psi}_{k}$. Then we have

\begin{equation}
\mathbf{y}^{T}L^{+}\mathbf{y}=\sum^{V-1}_{k=1}\frac{d^{2}_{k}}{\mu_{k}},\;(\mathbf{y}^{T}\mathbf{x})^{2}=\left(\sum^{V-1}_{k=1}c_{k}d_{k}\right)^{2},\;\mathbf{x}^{T}L\mathbf{x}=\sum^{V-1}_{k=1}\mu_{k}c^{2}_{k}.\label{eq_dec}
\end{equation}
Next, consider two vectors
$\mathbf{u}=(c_{1}\sqrt{\mu_{1}},\dots,c_{V-1}\sqrt{\mu_{V-1}})^{T}$
and
$\mathbf{v}=(d_{1}/\sqrt{\mu_{1}},\dots,d_{V-1}/\sqrt{\mu_{V-1}})^{T}$
and apply the classical Cauchy--Bunyakovsky inequality. We then obtain:
\[
(\mathbf{u}^{T}\mathbf{v})^{2}\le\|\mathbf{u}\|^{2}\|\mathbf{v}\|^{2}\implies\left(\sum^{V-1}_{k=1}c_{k}d_{k}\right)^{2}\le\left(\sum^{V-1}_{k=1}\mu_{k}c^{2}_{k}\right)\left(\sum^{V-1}_{k=1}\frac{d^{2}_{k}}{\mu_{k}}\right).
\]
Substituting here expressions (\ref{eq_dec}), we obtain
\[
(\mathbf{y}^{T}\mathbf{x})^{2}\le(\mathbf{x}^{T}L\mathbf{x})(\mathbf{y}^{T}L^{+}\mathbf{y}).
\]
Since $\mathbf{x}\in\mathrm{Im}(L)$ and $\mathbf{x}\neq\mathbf{0}$,
then $\mathbf{x}^{T}L\mathbf{x}>0$ and we arrive at the upper bound for
$\frac{(\mathbf{y}^{T}\mathbf{x})^{2}}{\mathbf{x}^{T}L\mathbf{x}}$:
\begin{equation}
\frac{(\mathbf{y}^{T}\mathbf{x})^{2}}{\mathbf{x}^{T}L\mathbf{x}}\le\mathbf{y}^{T}L^{+}\mathbf{y}.\label{eq_up}
\end{equation}

Let us prove that the found upper bound (\ref{eq_up}) is the exact
maximum. To do this, it suffices to indicate a value
$\mathbf{x}=\mathbf{x}^{*}$ at which this maximum is attained. Consider
the vector $\mathbf{x}^{*}=L^{+}\mathbf{y}\in\mathrm{Im}(L)$. Since
$\mathbf{y}\in\mathrm{Im}(L)$ and $\mathbf{y}\neq\mathbf{0}$, the
vector $\mathbf{x}^{*}\neq\mathbf{0}$ and, by symmetry of $L$,
$\mathbf{x}^{*}\in\mathrm{Im}(L^{+})=\mathrm{Im}(L)$. Computing the
value of the left-hand side of (\ref{eq_up}) on the vector
$\mathbf{x}^{*}$ using the relations
$(L^{+})^{T}=L^{+}$, $L^{+}LL^{+}=L^{+}$, it is easy to obtain
\[
\frac{(\mathbf{y}^{T}\mathbf{x}^{*})^{2}}{(\mathbf{x}^{*})^{T}L\mathbf{x}^{*}}=\frac{(\mathbf{y}^{T}L^{+}\mathbf{y})^{2}}{\mathbf{y}^{T}L^{+}\mathbf{y}}=\mathbf{y}^{T}L^{+}\mathbf{y}.
\]
Thus, the upper bound is attained on the vector
$\mathbf{x}^{*}=L^{+}\mathbf{y}$, which proves equality
(\ref{eq_var_principle}).

Next, in (\ref{eq_var_principle}) set $x_{i}=\sqrt{w_{i}}h_{i}$. The
condition $\mathbf{x}\in\mathrm{Im}(L)$ is equivalent to
$\langle\mathbf{x},\boldsymbol{\psi}_{0}\rangle=0$ or
$\sum_{i}w_{i}h_{i}=\langle\mathbf{w},\mathbf{h}\rangle=0$. The
numerator in (\ref{eq_var_principle}) transforms as follows:
\begin{equation}
\left(\mathbf{y}^{T}\mathbf{x}\right)^{2}=\left(\sum_{i\in\mathcal{B}_{a}}\frac{w_{i}}{W_{a}}h_{i}-\sum_{i\in\mathcal{B}_{b}}\frac{w_{i}}{W_{b}}h_{i}\right)^{2}=\left(\langle\mathbf{p}_{a},\mathbf{h}\rangle-\langle\mathbf{p}_{b},\mathbf{h}\rangle\right)^{2}.\label{eq_yx2}
\end{equation}
For the denominator, we use (\ref{eq_negative_semidef}) from Section
\ref{sec_kramers}:
\begin{equation}
\mathbf{x}^{T}L\mathbf{x}=\mathbf{x}^{T}(-S)\mathbf{x}=\sum_{\{i,j\}\in\mathcal{E}}S_{ij}\left(\sqrt[4]{\frac{w_{j}}{w_{i}}}x_{i}-\sqrt[4]{\frac{w_{i}}{w_{j}}}x_{j}\right)^{2}.\label{eq_xLx_pre}
\end{equation}
Substituting $x_{i}=\sqrt{w_{i}}h_{i}$ into (\ref{eq_xLx_pre}), taking
into account (\ref{eq_S_offdiag}), we find:
\begin{equation}
\mathbf{x}^{T}L\mathbf{x}=\sum_{\{i,j\}\in\mathcal{E}}\exp\!\bigl(-\beta\max(F_{i},F_{j})\bigr)(h_{i}-h_{j})^{2}=\mathcal{E}(\mathbf{h},\mathbf{h}).\label{eq_xLx}
\end{equation}
Substituting expressions (\ref{eq_yx2}) and (\ref{eq_xLx}) into
(\ref{eq_var_principle}) completes the proof of Lemma 1.

\textbf{Lemma 2 (Topological properties of the single-linkage distance).}
Let $\rho=\rho^{(0)}(\mathcal{B}_{a},\mathcal{B}_{b})$. Consider the
subgraph $G_{<\rho}$ containing all vertices $\mathcal{M}$ and only
those edges $\{i,j\}\in\mathcal{E}$ for which
$U_{ij}=\max(F_{i},F_{j})<\rho$. Then the sets $\mathcal{B}_{a}$ and
$\mathcal{B}_{b}$ belong to different connected components of subgraph
$G_{<\rho}$.

\textbf{Proof of Lemma 2.} Assume the contrary: let $\mathcal{B}_{a}$
and $\mathcal{B}_{b}$ lie in one connected component of $G_{<\rho}$.
Then there exists a path $\gamma$ between some vertices
$p\in\mathcal{B}_{a}$ and $q\in\mathcal{B}_{b}$, all edges of which
have weight strictly less than $\rho$. Consequently, the height of this
path $h(\gamma)<\rho$. By definition of the single-linkage distance,
$\rho^{(0)}(\mathcal{B}_{a},\mathcal{B}_{b})\le h(\gamma)<\rho$, which
contradicts the equality
$\rho^{(0)}(\mathcal{B}_{a},\mathcal{B}_{b})=\rho$. Lemma 2 is proved.

\textbf{Definition of the asymptotic constant $\mathcal{R}^{*}_{ab}$.}
The constant $\mathcal{R}^{*}_{ab}$ depends on the pair of basins
$(\mathcal{B}_{a},\mathcal{B}_{b})$ and is defined as follows. Let
$i_{a}=\arg\min_{i\in\mathcal{B}_{a}}F_{i}$ be the vertex with minimum
free energy in basin $\mathcal{B}_{a}$, and
$i_{b}=\arg\min_{j\in\mathcal{B}_{b}}F_{j}$ be the vertex with minimum
free energy in basin $\mathcal{B}_{b}$. Consider the subgraph
$G_{<\rho}$ of graph $G$, containing all vertices $\mathcal{M}$ and
only those edges $\{i,j\}\in\mathcal{E}$ for which
$U_{ij}=\max(F_{i},F_{j})<\rho$, where
$\rho=\rho^{(0)}(\mathcal{B}_{a},\mathcal{B}_{b})$. Let
$\mathcal{V}^{*}=\{\mathcal{C}_{1},\dots,\mathcal{C}_{M}\}$ be the set
of connected components of subgraph $G_{<\rho}$. Denote by
$\mathcal{C}_{a}\in\mathcal{V}^{*}$ the connected component of subgraph
$G_{<\rho}$ containing the vertex $i_{a}\in\mathcal{B}_{a}$, and by
$\mathcal{C}_{b}\in\mathcal{V}^{*}$ the component containing
$i_{b}\in\mathcal{B}_{b}$. Since, according to Lemma 2, the basins
$\mathcal{B}_{a}$ and $\mathcal{B}_{b}$ lie entirely in different
connected components of subgraph $G_{<\rho}$, the vertices $i_{a}$ and
$i_{b}$ cannot be connected by a path in this subgraph, and,
consequently, $\mathcal{C}_{a}\neq\mathcal{C}_{b}$.

It is important to note that the connected component $\mathcal{C}_{a}$,
generally speaking, does not coincide with the basin of attraction
$\mathcal{B}_{a}$. The set $\mathcal{B}_{a}$ is identified by the
gradient descent algorithm (moving strictly downhill in energy towards
the point $i_{a}$), whereas $\mathcal{C}_{a}$ is determined by
connectivity through edges with weights $U_{ij}<\rho$, which allows
movement both downhill and uphill, provided the threshold $\rho$ is not
exceeded. As a result, two types of mismatches are possible:

(1) vertices with high energies $F_{j}\ge\rho$, from which gradient
descent leads to $i_{a}$, belong to $\mathcal{B}_{a}$ but are isolated
in $G_{<\rho}$ and are not included in $\mathcal{C}_{a}$;

(2) neighboring local minima $i_{c}\neq i_{a}$, forming their own
basins $\mathcal{B}_{c}\neq\mathcal{B}_{a}$, may belong to
$\mathcal{C}_{a}$ if the barrier separating them from $i_{a}$ is
strictly less than $\rho$.

Define the contracted graph $\mathcal{G}^{*}=(\mathcal{V}^{*},\mathcal{E}^{*})$,
the vertex set of which $\mathcal{V}^{*}$ is formed by the connected
components of subgraph $G_{<\rho}$. The edge set $\mathcal{E}^{*}$ is
defined as follows: two vertices
$\mathcal{C}_{u},\mathcal{C}_{v}\in\mathcal{V}^{*}$ are connected by
an edge $e=(\mathcal{C}_{u},\mathcal{C}_{v})\in\mathcal{E}^{*}$ if and
only if in the original graph $G$ there exists at least one edge
$\{i,j\}\in\mathcal{E}$ such that $i\in\mathcal{C}_{u}$,
$j\in\mathcal{C}_{v}$ and the weight of this edge
$U_{ij}=\max(F_{i},F_{j})$ is exactly equal to $\rho$. To each edge
$e\in\mathcal{E}^{*}$ we assign an integer weight $N_{e}$, equal to the
number of such minimax edges between components $\mathcal{C}_{u}$ and
$\mathcal{C}_{v}$ in the original graph $G$.

Let $M=|\mathcal{V}^{*}|$ be the number of vertices of graph
$\mathcal{G}^{*}$. Consider vectors $\mathbf{g}$ whose components are
indexed by the elements of the set $\mathcal{V}^{*}$. In other words,
each such vector is given by a set of $M$ real numbers
$\{g_{\mathcal{C}}\}_{\mathcal{C}\in\mathcal{V}^{*}}$, where the index
$\mathcal{C}$ runs over all vertices of graph $\mathcal{G}^{*}$. In
this notation, the entry $g_{\mathcal{C}_{u}}$ means the component of
the vector corresponding to the vertex
$\mathcal{C}_{u}\in\mathcal{V}^{*}$.

Define the quadratic form $Q(\mathbf{g})$, associating with each vector
$\mathbf{g}$ a real number:
\begin{equation}
Q(\mathbf{g})=\sum_{e=(\mathcal{C}_{u},\mathcal{C}_{v})\in\mathcal{E}^{*}}N_{e}\bigl(g_{\mathcal{C}_{u}}-g_{\mathcal{C}_{v}}\bigr)^{2}.\label{eq_Q_form}
\end{equation}
We define the constant $\mathcal{R}^{*}_{ab}$ as the exact supremum of
the ratio of the squared difference of vector components corresponding
to the distinguished vertices to the value of the quadratic form:
\begin{equation}
\mathcal{R}^{*}_{ab}=\sup_{\mathbf{g}:\,g_{\mathcal{C}_{a}}\neq g_{\mathcal{C}_{b}}}\frac{\bigl(g_{\mathcal{C}_{a}}-g_{\mathcal{C}_{b}}\bigr)^{2}}{Q(\mathbf{g})}.\label{eq_R_star_def}
\end{equation}

We show that this supremum is attained (i.e., is a maximum), and the
quantity $\mathcal{R}^{*}_{ab}$ is a positive finite constant.

The graph $\mathcal{G}^{*}$, generally speaking, is not necessarily
connected. However, since $\rho$ by definition is the minimax barrier
between basins $\mathcal{B}_{a}$ and $\mathcal{B}_{b}$, in the graph
$G$ there exists a path between them whose height equals $\rho$. This
guarantees that in graph $\mathcal{G}^{*}$, the vertices
$\mathcal{C}_{a}$ and $\mathcal{C}_{b}$ belong to one and the same
connected component, which we denote by $\mathcal{K}$.

Maximization of the fraction in (\ref{eq_R_star_def}) is equivalent to
minimizing the denominator $Q(\mathbf{g})$ under the condition
$g_{\mathcal{C}_{a}}-g_{\mathcal{C}_{b}}=1$ (by virtue of the
homogeneity of the functional). The contribution to $Q(\mathbf{g})$ from
connected components that do not intersect $\mathcal{K}$ is non-negative
and does not affect the constraints; therefore, to minimize, it is
sufficient to set $g_{\mathcal{C}}$ constant inside each such component.
Hence, the problem reduces to minimizing the quadratic form
$Q_{\mathcal{K}}(\mathbf{g})=\sum_{e\in\mathcal{E}(\mathcal{K})}N_{e}(g_{u}-g_{v})^{2}$
on the closed affine subspace
\[
W_{\mathcal{K}}=\{\mathbf{g}\in\mathbb{R}^{|\mathcal{K}|}\mid g_{\mathcal{C}_{a}}=1,\,g_{\mathcal{C}_{b}}=0\}.
\]

Let us investigate the properties of the form $Q_{\mathcal{K}}$. By
definition, it is non-negative. Its kernel (the set of vectors on which
$Q_{\mathcal{K}}(\mathbf{g})=0$) consists of vectors for which
$g_{u}=g_{v}$ on all edges of the connected component $\mathcal{K}$.
Consequently, the kernel is formed by vectors with identical components
(proportional to the vector of ones
$\mathbf{1}=(1,1,\dots,1)^{T}$).

According to the standard orthogonal decomposition in Euclidean space,
any vector $\mathbf{g}\in\mathbb{R}^{|\mathcal{K}|}$ can be uniquely
represented as the sum of a vector proportional to $\mathbf{1}$ and a
vector orthogonal to it:
\[
\mathbf{g}=c\mathbf{1}+\mathbf{x},
\]
where the constant $c\in\mathbb{R}$, and the vector $\mathbf{x}$
belongs to the subspace
$V_{0}=\{\mathbf{x}\in\mathbb{R}^{|\mathcal{K}|}\mid\sum_{i\in\mathcal{K}}x_{i}=0\}$
(vectors with zero sum of components).

Since adding a constant to all components of a vector does not change
the differences $(g_{u}-g_{v})$, the form is invariant under a shift by
$c\mathbf{1}$: $Q_{\mathcal{K}}(\mathbf{g})=Q_{\mathcal{K}}(\mathbf{x})$.
Consider the restriction of $Q_{\mathcal{K}}$ to the finite-dimensional
subspace $V_{0}$. If for some $\mathbf{x}\in V_{0}$ we have
$Q_{\mathcal{K}}(\mathbf{x})=0$, then $\mathbf{x}$ must belong to the
kernel of the form, i.e., have identical components. But the only vector
with identical components whose sum equals zero is the zero vector.
Hence, on the subspace $V_{0}$ the kernel is trivial. Any non-negative
quadratic form with a trivial kernel on a finite-dimensional space is
strictly positive definite: there exists a constant $\lambda>0$ (the
smallest eigenvalue of the corresponding matrix on $V_{0}$) such that
$Q_{\mathcal{K}}(\mathbf{x})\ge\lambda\|\mathbf{x}\|^{2}$ for all
$\mathbf{x}\in V_{0}$.

Let us prove the coercivity of $Q_{\mathcal{K}}$ on $W_{\mathcal{K}}$,
i.e., that $Q_{\mathcal{K}}(\mathbf{g})\to\infty$ as
$\|\mathbf{g}\|\to\infty$. From the constraints
$g_{\mathcal{C}_{a}}=1$ and $g_{\mathcal{C}_{b}}=0$, it follows that
$x_{\mathcal{C}_{a}}=1-c$ and $x_{\mathcal{C}_{b}}=-c$. The squared
norm of the vector $\mathbf{x}$ is bounded below by the sum of squares
of these two components:
\[
\|\mathbf{x}\|^{2}\ge(1-c)^{2}+c^{2}\ge\frac{1}{2}.
\]
Moreover, by orthogonality of $\mathbf{1}$ and $\mathbf{x}$, by the
Pythagorean theorem,
$\|\mathbf{g}\|^{2}=\|c\mathbf{1}\|^{2}+\|\mathbf{x}\|^{2}=c^{2}|\mathcal{K}|+\|\mathbf{x}\|^{2}$.
From this it is obvious that the tendency $\|\mathbf{g}\|\to\infty$
entails $\|\mathbf{x}\|\to\infty$ (if $\|\mathbf{x}\|$ were bounded,
then $c^{2}\le\|\mathbf{x}\|^{2}$ would also be bounded, and hence
$\|\mathbf{g}\|$ would be bounded). Consequently, as
$\|\mathbf{g}\|\to\infty$, the quantity
$Q_{\mathcal{K}}(\mathbf{g})=Q_{\mathcal{K}}(\mathbf{x})\ge\lambda\|\mathbf{x}\|^{2}$
also tends to infinity.

Thus, we are minimizing a continuous coercive function on a closed set.
By the Weierstrass theorem, the minimum is attained on some vector
$\mathbf{g}^{*}\in W_{\mathcal{K}}$. The minimum value is strictly
positive, since
$Q_{\mathcal{K}}(\mathbf{g}^{*})\ge\lambda\|\mathbf{x}^{*}\|^{2}\ge\lambda/2>0$.
Consequently, the supremum in (\ref{eq_R_star_def}) is attained, and
the constant $\mathcal{R}^{*}_{ab}=1/Q_{\mathcal{K}}(\mathbf{g}^{*})$
is a finite positive quantity, depending only on the topology of the
minimax edges and the structure of the connected components of
$G_{<\rho}$.

\textbf{Proof of the Theorem.} We prove the exact asymptotic expansion
$D^{2}_{\beta}=\exp(\beta\rho)\left[\mathcal{R}^{*}_{ab}+O\!\left(e^{-\beta\delta}\right)\right]$,
from which the logarithmic asymptotics is obtained as a direct
consequence. We divide the proof into 3 steps.

Step 1 (finding a precise lower bound with extraction of the leading
term). According to Lemma 1, the square of the kinetic metric
$D^{2}_{\beta}$ admits the variational representation in the form
(\ref{eq_Lemma_1}). Since the global maximum of the functional is not
less than its value at any concrete admissible point, to obtain a lower
bound for $D^{2}_{\beta}$ it suffices to compute the value of this
fraction on some specially chosen vector $\mathbf{h}^{\ast}$ satisfying
the orthogonality condition
$\langle\mathbf{w},\mathbf{h}^{\ast}\rangle=0$. In variational
calculus, such a vector, substituted into the functional to obtain an
estimate of the extremum, is customarily called a trial vector.

Our task is to construct a trial vector that asymptotically accurately
reflects the structure of the minimax barriers between basins
$\mathcal{B}_{a}$ and $\mathcal{B}_{b}$. To this end, we will rely on
the optimal vector $\mathbf{g}^{*}$, defined on the contracted graph
$\mathcal{G}^{*}$ in the definition of the asymptotic constant
$\mathcal{R}^{*}_{ab}$. However, before proceeding to the construction
of the trial vector and the estimation of the functional, we prove two
auxiliary topological statements about the structure of basins and
connected components of subgraph $G_{<\rho}$.

Recall that $i_{a}=\arg\min_{i\in\mathcal{B}_{a}}F_{i}$ is the vertex
with minimum energy in basin $\mathcal{B}_{a}$, and
$\mathcal{C}_{a}\in\mathcal{V}^{*}$ is the connected component of
subgraph $G_{<\rho}$ containing the vertex $i_{a}$. The vertex $i_{b}$
and component $\mathcal{C}_{b}$ for basin $\mathcal{B}_{b}$ are defined
analogously.

Let us first prove two statements.

Statement 1: for any $a\neq b$,
$\mathcal{B}_{b}\cap\mathcal{C}_{a}=\emptyset$. To prove this
statement, suppose the contrary: let there exist a vertex
$v\in\mathcal{B}_{b}\cap\mathcal{C}_{a}$. Since
$i_{a}\in\mathcal{C}_{a}$ and $v\in\mathcal{C}_{a}$, and
$\mathcal{C}_{a}$ is a connected component of subgraph $G_{<\rho}$,
between vertices $i_{a}$ and $v$ there exists a path lying entirely in
$G_{<\rho}$. However, $i_{a}\in\mathcal{B}_{a}$ and
$v\in\mathcal{B}_{b}$, and, consequently, this path connects basin
$\mathcal{B}_{a}$ with basin $\mathcal{B}_{b}$ in subgraph $G_{<\rho}$.
This directly contradicts the statement of Lemma 2, according to which
basins $\mathcal{B}_{a}$ and $\mathcal{B}_{b}$ lie in different
connected components of $G_{<\rho}$. The obtained contradiction proves
that $\mathcal{B}_{b}\cap\mathcal{C}_{a}=\emptyset$.

Statement 2: for any vertex $j\in\mathcal{B}_{a}\setminus\mathcal{C}_{a}$,
we have $F_{j}\ge\rho$. To prove the statement, suppose that
$F_{j}<\rho$. By definition of a basin of attraction, there exists a
gradient descent path from $j$ to $i_{a}$, along which the free energy
strictly decreases. Since the initial energy $F_{j}<\rho$ and the
energy decreases, all vertices and edges on this path have weight
strictly less than $\rho$. Consequently, this entire path lies in
subgraph $G_{<\rho}$, and vertex $j$ must belong to the same connected
component as $i_{a}$ (i.e., $j\in\mathcal{C}_{a}$), which contradicts
the condition $j\notin\mathcal{C}_{a}$. Thus, $F_{j}\ge\rho$, and,
accordingly, the Boltzmann weight of such vertices satisfies the
estimate $w_{j}\le\exp(-\beta\rho)$. Note that by symmetry of the
definitions, the analogous property $F_{j}\ge\rho$ also holds for all
vertices $j\in\mathcal{B}_{b}\setminus\mathcal{C}_{b}$.

Now we proceed to the construction of the trial vector. As was shown in
the justification of the definition of the constant
$\mathcal{R}^{*}_{ab}$, the supremum in formula (\ref{eq_R_star_def})
is attained on some optimal vector $\mathbf{g}^{*}$, defined on the
vertices of the contracted graph $\mathcal{G}^{*}$. By virtue of the
invariance of the maximized fraction under a shift by a constant, we can
fix the normalization:
\begin{equation}
g^{*}_{\mathcal{C}_{a}}=1,\;g^{*}_{\mathcal{C}_{b}}=0,\;Q(\mathbf{g}^{*})=\frac{1}{\mathcal{R}^{*}_{ab}}.\label{eq_g_star_props_new}
\end{equation}
We construct a special trial vector $\mathbf{h}^{*}\in\mathbb{R}^{V}$
for the original graph, relying on the vector $\mathbf{g}^{*}$. For
each vertex $i\in\mathcal{M}$, we find the connected component
$\mathcal{C}(i)\in\mathcal{V}^{*}$ of subgraph $G_{<\rho}$ to which it
belongs (if the vertex $i$ is isolated in $G_{<\rho}$, then
$\mathcal{C}(i)=\{i\}$), and assign to it the corresponding value from
the vector $\mathbf{g}^{*}$:
\begin{equation}
h^{*}_{i}=g^{*}_{\mathcal{C}(i)},\;i\in\mathcal{M}.\label{eq_h_star_def_new}
\end{equation}
By this definition, the values $h^{*}_{i}$ remain constant within each
connected component of $G_{<\rho}$. The vector $\mathbf{h}^{*}$ may not
satisfy the orthogonality condition
$\langle\mathbf{w},\mathbf{h}^{*}\rangle=0$ required in Lemma 1.
However, introducing the shift
$\tilde{\mathbf{h}}^{*}=\mathbf{h}^{*}-c\mathbf{1}$, where
$c=\langle\mathbf{w},\mathbf{h}^{*}\rangle/\langle\mathbf{w},\mathbf{1}\rangle$,
we obtain an admissible trial vector. The numerator of the variational
functional is invariant under such a shift, since
$\langle\mathbf{p}_{a}-\mathbf{p}_{b},\mathbf{1}\rangle=1-1=0$, and
the denominator (the quadratic form $\mathcal{E}$) depends only on the
differences of values on neighboring vertices and is also invariant:
$\mathcal{E}(\tilde{\mathbf{h}}^{*},\tilde{\mathbf{h}}^{*})=\mathcal{E}(\mathbf{h}^{*},\mathbf{h}^{*})$.
Therefore, we can compute the functional directly on the vector
$\mathbf{h}^{*}$.

Let us estimate the numerator of the variational functional for the
trial vector $\mathbf{h}^{*}$. We split the sum
$\langle\mathbf{p}_{a},\mathbf{h}^{*}\rangle=\sum_{j\in\mathcal{B}_{a}}p_{a,j}h^{*}_{j}$
into two parts: over the set $\mathcal{B}_{a}\cap\mathcal{C}_{a}$ and
over $\mathcal{B}_{a}\setminus\mathcal{C}_{a}$. For any vertex
$j\in\mathcal{B}_{a}\cap\mathcal{C}_{a}$, we have
$h^{*}_{j}=g^{*}_{\mathcal{C}_{a}}=1$. For vertices
$j\in\mathcal{B}_{a}\setminus\mathcal{C}_{a}$, according to Statement
2, we have $F_{j}\ge\rho$. We show that their total contribution to the
average value is negligible as $\beta\to\infty$. First, estimate the
total probability (statistical weight) of this set:
\[
\sum_{j\in\mathcal{B}_{a}\setminus\mathcal{C}_{a}}p_{a,j}=\frac{\sum_{j\in\mathcal{B}_{a}\setminus\mathcal{C}_{a}}w_{j}}{W_{a}}\le\frac{V\exp(-\beta\rho)}{\exp(-\beta F_{i_{a}})}=V\exp\bigl(-\beta(\rho-F_{i_{a}})\bigr).
\]
Since the minimax barrier is strictly higher than the basin bottom
energy ($\rho>F_{i_{a}}$), the difference $\Delta=\rho-F_{i_{a}}$ is
positive, and the sum of probabilities tends exponentially to zero,
i.e., equals $o(1)$. Second, the values $h^{*}_{j}$ are bounded by the
constant $C_{\max}=\max_{C\in\mathcal{V}^{*}}|g^{*}_{C}|$. Since the
graph $\mathcal{G}^{*}$ and its edge weights are independent of $\beta$,
the constant $C_{\max}$ is finite and independent of $\beta$.
Consequently, the modulus of the contribution of these vertices is
bounded by the product of the constant and the exponentially small
probability:
\[
\left|\sum_{j\in\mathcal{B}_{a}\setminus\mathcal{C}_{a}}p_{a,j}h^{*}_{j}\right|\le C_{\max}\sum_{j\in\mathcal{B}_{a}\setminus\mathcal{C}_{a}}p_{a,j}\le C_{\max}V\exp(-\beta\Delta)=o(1).
\]
Thus, $\langle\mathbf{p}_{a},\mathbf{h}^{*}\rangle=1+o(1)$.

Similarly, for basin $\mathcal{B}_{b}$, the main statistical weight is
concentrated in the intersection
$\mathcal{B}_{b}\cap\mathcal{C}_{b}$, where
$h^{*}_{j}=g^{*}_{\mathcal{C}_{b}}=0$. For the same reason as in
Statement 2, for any vertex
$j\in\mathcal{B}_{b}\setminus\mathcal{C}_{b}$, we have $F_{j}\ge\rho$,
and their total contribution is bounded by
$C_{\max}V\exp(-\beta(\rho-F_{i_{b}}))=o(1)$. Consequently, taking into
account the normalization (\ref{eq_g_star_props_new}), we obtain:
\[
\langle\mathbf{p}_{a},\mathbf{h}^{*}\rangle=1+o(1),\;\langle\mathbf{p}_{b},\mathbf{h}^{*}\rangle=0+o(1).
\]
Thus, the numerator of the variational functional takes the form:
\[
\bigl(\langle\mathbf{p}_{a},\mathbf{h}^{*}\rangle-\langle\mathbf{p}_{b},\mathbf{h}^{*}\rangle\bigr)^{2}=(1+o(1))^{2}=1+o(1).
\]

Now we estimate the denominator
$\mathcal{E}(\mathbf{h}^{*},\mathbf{h}^{*})$. The quadratic form
$\mathcal{E}$ is the sum over all edges of the graph:
\[
\mathcal{E}(\mathbf{h}^{*},\mathbf{h}^{*})=\sum_{\{i,j\}\in\mathcal{E}}\exp\bigl(-\beta\max(F_{i},F_{j})\bigr)(h^{*}_{i}-h^{*}_{j})^{2}.
\]
For any edges lying entirely within one connected component of
$G_{<\rho}$, the difference $(h^{*}_{i}-h^{*}_{j})^{2}$ is identically
zero. Consequently, a non-zero contribution is made only by edges
connecting different components. We split these edges into two groups:

1. Minimax edges, for which $U_{ij}=\max(F_{i},F_{j})=\rho$. The weight
of such edges is $\exp(-\beta\rho)$. The summation of the squared
differences $(h^{*}_{i}-h^{*}_{j})^{2}$ over all minimax edges exactly
reproduces the quadratic form $Q(\mathbf{g}^{*})$, defined on the
contracted graph $\mathcal{G}^{*}$.

2. Suboptimal edges, for which $U_{ij}\ge\rho+\delta$. The weight of
such edges does not exceed $\exp(-\beta(\rho+\delta))$. Since the
differences $(h^{*}_{i}-h^{*}_{j})^{2}$ are bounded by the constant
$4C^{2}_{\max}$, and the number of edges is finite, their total
contribution has order $O(e^{-\beta(\rho+\delta)})$.

Combining these estimates, we obtain:
\[
\mathcal{E}(\mathbf{h}^{*},\mathbf{h}^{*})=\exp(-\beta\rho)\cdot Q(\mathbf{g}^{*})+O(e^{-\beta(\rho+\delta)}).
\]
Substituting here the property of the optimal vector
$Q(\mathbf{g}^{*})=1/\mathcal{R}^{*}_{ab}$ from
(\ref{eq_g_star_props_new}), we find:
\[
\mathcal{E}(\mathbf{h}^{*},\mathbf{h}^{*})=\frac{\exp(-\beta\rho)}{\mathcal{R}^{*}_{ab}}+O(e^{-\beta(\rho+\delta)}).
\]

According to Lemma 1, the quantity $D^{2}_{\beta}$ is exactly equal to
the maximum of the variational functional
$J(\mathbf{h})=\frac{\bigl(\langle\mathbf{p}_{a},\mathbf{h}\rangle-\langle\mathbf{p}_{b},\mathbf{h}\rangle\bigr)^{2}}{\mathcal{E}(\mathbf{h},\mathbf{h})}$
over all admissible vectors $\mathbf{h}$. Since the global maximum is
not less than the value of the function at any concrete point, for our
admissible trial vector $\tilde{\mathbf{h}}^{*}$ the inequality holds:
\begin{equation}
D^{2}_{\beta}\ge J(\tilde{\mathbf{h}}^{*})=J(\mathbf{h}^{*})=\frac{\bigl(\langle\mathbf{p}_{a},\mathbf{h}^{*}\rangle-\langle\mathbf{p}_{b},\mathbf{h}^{*}\rangle\bigr)^{2}}{\mathcal{E}(\mathbf{h}^{*},\mathbf{h}^{*})}.\label{eq_var_ineq_new}
\end{equation}
Substituting the found asymptotic expressions for the numerator and
denominator, we obtain:
\[
D^{2}_{\beta}\ge\frac{1+o(1)}{\frac{\exp(-\beta\rho)}{\mathcal{R}^{*}_{ab}}+O(e^{-\beta(\rho+\delta)})}=\exp(\beta\rho)\left[\mathcal{R}^{*}_{ab}+O\!\left(e^{-\beta\delta}\right)\right].
\]
From which follows the desired exact lower bound:
\begin{equation}
D^{2}_{\beta}\ge\exp(\beta\rho)\left[\mathcal{R}^{*}_{ab}+O\!\left(e^{-\beta\delta}\right)\right].\label{eq_lower_bound_precise}
\end{equation}

Step 2 (finding an upper bound with extraction of the leading term).
Let $\mathbf{h}\in\mathbb{R}^{V}$ be an arbitrary vector satisfying the
condition $\langle\mathbf{w},\mathbf{h}\rangle=0$. Denote
$\Delta=\langle\mathbf{p}_{a},\mathbf{h}\rangle-\langle\mathbf{p}_{b},\mathbf{h}\rangle$.
For each connected component $\mathcal{C}\in\mathcal{V}^{*}$ of
subgraph $G_{<\rho}$, denote by
$v_{\mathcal{C}}=\arg\min_{v\in\mathcal{C}}F_{v}$ the vertex with
minimum energy in this component, and define the vector $\mathbf{g}$ by
setting $g_{\mathcal{C}}=h_{v_{\mathcal{C}}}$.

We show that the values of $\mathbf{h}$ inside each component
$\mathcal{C}$ differ from $g_{\mathcal{C}}$ by an exponentially small
quantity. Indeed, any two vertices $i,j\in\mathcal{C}$ are connected by
a path $\gamma=(u_{0},u_{1},\dots,u_{L})$ of length $L\le V$, all of
whose edges have weight
$U_{u_{k}u_{k+1}}\le\rho_{\mathcal{C}}<\rho$, where
$\rho_{\mathcal{C}}=\max_{\{u,v\}\in\mathcal{E}(\mathcal{C})}U_{uv}$.
The difference of the function values at the ends of the path can be
written as the sum of successive differences along the edges:
\begin{equation}
h_{i}-h_{j}=\sum^{L-1}_{k=0}(h_{u_{k}}-h_{u_{k+1}}),\label{eq_sum_h}
\end{equation}
where all intermediate terms cancel. Represent each term in the sum
(\ref{eq_sum_h}) as the product
\[
h_{u_{k}}-h_{u_{k+1}}=\frac{1}{\sqrt{C_{u_{k}u_{k+1}}}}\cdot\sqrt{C_{u_{k}u_{k+1}}}(h_{u_{k}}-h_{u_{k+1}}),
\]
where $C_{ij}$ are defined in (\ref{eq_Lemma_1}), and apply the
standard Cauchy--Bunyakovsky inequality
$\left(\sum a_{k}b_{k}\right)^{2}\le\left(\sum a^{2}_{k}\right)\left(\sum b^{2}_{k}\right)$,
setting $a_{k}=1/\sqrt{C_{u_{k}u_{k+1}}}$ and
$b_{k}=\sqrt{C_{u_{k}u_{k+1}}}(h_{u_{k}}-h_{u_{k+1}})$:
\[
|h_{i}-h_{j}|^{2}\le\left(\sum^{L-1}_{k=0}\frac{1}{C_{u_{k}u_{k+1}}}\right)\left(\sum^{L-1}_{k=0}C_{u_{k}u_{k+1}}(h_{u_{k}}-h_{u_{k+1}})^{2}\right).
\]
Since all edges of the path lie in component $\mathcal{C}$, for them
$U_{u_{k}u_{k+1}}\le\rho_{\mathcal{C}}$, whence
$C_{u_{k}u_{k+1}}=\exp(-\beta U_{u_{k}u_{k+1}})\ge\exp(-\beta\rho_{\mathcal{C}})$.
Consequently, the first bracket does not exceed
$L\exp(\beta\rho_{\mathcal{C}})\le V\exp(\beta\rho_{\mathcal{C}})$.
The second bracket is part of the full quadratic form and is bounded
above by $\mathcal{E}(\mathbf{h},\mathbf{h})$. Thus:
\[
|h_{i}-h_{j}|\le\sqrt{V\exp(\beta\rho_{\mathcal{C}})\mathcal{E}(\mathbf{h},\mathbf{h})}=o\!\left(\exp(\beta\rho/2)\sqrt{\mathcal{E}(\mathbf{h},\mathbf{h})}\right),
\]
since $\rho_{\mathcal{C}}<\rho$. Hence, for any vertex $i$ in component
$\mathcal{C}$, we have
$h_{i}=g_{\mathcal{C}}+o(\exp(\beta\rho/2)\sqrt{\mathcal{E}(\mathbf{h},\mathbf{h})})$.

For basin $\mathcal{B}_{a}$, we split the sum
$\langle\mathbf{p}_{a},\mathbf{h}\rangle=\sum_{j\in\mathcal{B}_{a}}(p_{a})_{j}h_{j}$
into two sums -- the sum over $\mathcal{B}_{a}\cap\mathcal{C}_{a}$ and
the sum over $\mathcal{B}_{a}\setminus\mathcal{C}_{a}$. For
$j\in\mathcal{B}_{a}\cap\mathcal{C}_{a}$, we have
$h_{j}=g_{\mathcal{C}_{a}}+o(\exp(\beta\rho/2)\sqrt{\mathcal{E}})$.
For $j\in\mathcal{B}_{a}\setminus\mathcal{C}_{a}$, as proved in Step
1, $F_{j}\ge\rho$, and their Boltzmann weight
$w_{j}\le\exp(-\beta\rho)$. The contribution of these vertices to the
average value is estimated as
\[
(p_{a})_{j}|h_{j}|\le\frac{\exp(-\beta F_{j})}{W_{a}}\cdot O(|h_{j}|)=o\!\left(\exp(\beta\rho/2)\sqrt{\mathcal{E}(\mathbf{h},\mathbf{h})}\right),
\]
since
$\sum_{j\in\mathcal{B}_{a}\setminus\mathcal{C}_{a}}(p_{a})_{j}\le V\exp(-\beta\rho)/W_{a}=o(1)$.
Consequently,
\[
\langle\mathbf{p}_{a},\mathbf{h}\rangle=g_{\mathcal{C}_{a}}+\varepsilon_{a},\;|\varepsilon_{a}|=o\!\left(\exp(\beta\rho/2)\sqrt{\mathcal{E}(\mathbf{h},\mathbf{h})}\right).
\]
Analogously, for $\mathcal{B}_{b}$ we have
$\langle\mathbf{p}_{b},\mathbf{h}\rangle=g_{\mathcal{C}_{b}}+\varepsilon_{b}$
with an exponentially suppressed error $\varepsilon_{b}$. Therefore,
\begin{equation}
\Delta^{2}=\bigl(g_{\mathcal{C}_{a}}-g_{\mathcal{C}_{b}}+\varepsilon_{a}-\varepsilon_{b}\bigr)^{2}\le\bigl(g_{\mathcal{C}_{a}}-g_{\mathcal{C}_{b}}\bigr)^{2}\cdot(1+o(1)).\label{eq_Delta}
\end{equation}

Now we estimate the quadratic form $\mathcal{E}(\mathbf{h},\mathbf{h})$
from below via $Q(\mathbf{g})$. We split the set of all edges of the
graph $\mathcal{E}$ into three disjoint subsets:

1. $E_{1}$ -- the subset of edges lying entirely within connected
components of subgraph $G_{<\rho}$.

2. $E_{2}$ -- the subset of edges between different components (for
which $U_{ij}\ge\rho+\delta$).

3. $E_{3}$ -- the subset of edges between different components (for
which $U_{ij}=\rho$).

Then the total sum can be identically rewritten as the sum over these
three subsets:
\[
\mathcal{E}(\mathbf{h},\mathbf{h})=\sum_{\{i,j\}\in E_{1}}C_{ij}(h_{i}-h_{j})^{2}+\sum_{\{i,j\}\in E_{2}}C_{ij}(h_{i}-h_{j})^{2}+\sum_{\{i,j\}\in E_{3}}C_{ij}(h_{i}-h_{j})^{2}.
\]
Since each term $C_{ij}(h_{i}-h_{j})^{2}$ is non-negative, the partial
sums over subsets $E_{1}$ and $E_{2}$ are also non-negative:
\[
\sum_{\{i,j\}\in E_{1}}C_{ij}(h_{i}-h_{j})^{2}\ge0,\;\sum_{\{i,j\}\in E_{2}}C_{ij}(h_{i}-h_{j})^{2}\ge0.
\]
Leaving only the sum over $E_{3}$, we obtain for
$\mathcal{E}(\mathbf{h},\mathbf{h})$ an estimate of the form
\begin{equation}
\mathcal{E}(\mathbf{h},\mathbf{h})\ge\sum_{\{i,j\}\in E_{3}}C_{ij}(h_{i}-h_{j})^{2}.\label{eq_E_h_ge}
\end{equation}
We next work only with the sum over edges $E_{3}$. For each such edge
$\{i,j\}\in E_{3}$, connecting components $\mathcal{C}_{u}$ and
$\mathcal{C}_{v}$, the weight is exactly
$C_{ij}=\exp(-\beta\rho)$. As was shown above, the values $h_{i}$ and
$h_{j}$ differ from $g_{\mathcal{C}_{u}}$ and $g_{\mathcal{C}_{v}}$
by an exponentially small quantity:
\[
h_{i}=g_{\mathcal{C}_{u}}+o\!\left(\exp(\beta\rho/2)\sqrt{\mathcal{E}(\mathbf{h},\mathbf{h})}\right),\;h_{j}=g_{\mathcal{C}_{v}}+o\!\left(\exp(\beta\rho/2)\sqrt{\mathcal{E}(\mathbf{h},\mathbf{h})}\right),
\]
consequently
\[
(h_{i}-h_{j})^{2}=\bigl(g_{\mathcal{C}_{u}}-g_{\mathcal{C}_{v}}+o\!\left(\exp(\beta\rho/2)\sqrt{\mathcal{E}}\right)\bigr)^{2}.
\]
Then, taking into account (\ref{eq_Q_form}), the estimate
(\ref{eq_E_h_ge}) can be written in the form
\begin{equation}
\mathcal{E}(\mathbf{h},\mathbf{h})\ge\exp(-\beta\rho)\sum_{e=(\mathcal{C}_{u},\mathcal{C}_{v})\in\mathcal{E}^{*}}N_{e}\bigl(g_{\mathcal{C}_{u}}-g_{\mathcal{C}_{v}}\bigr)^{2}\cdot(1-o(1))=\exp(-\beta\rho)\cdot Q(\mathbf{g})\cdot(1-o(1)),\label{eq_E_g_e}
\end{equation}
From the definition (\ref{eq_R_star_def}) of the constant
$\mathcal{R}^{*}_{ab}$, it follows that
\begin{equation}
\bigl(g_{\mathcal{C}_{a}}-g_{\mathcal{C}_{b}}\bigr)^{2}\le\mathcal{R}^{*}_{ab}\cdot Q(\mathbf{g}).\label{eq_g_R_Q}
\end{equation}
Combining (\ref{eq_g_R_Q}) with the estimates (\ref{eq_Delta}) for
$\Delta^{2}=\left(\langle\mathbf{p}_{a},\mathbf{h}\rangle-\langle\mathbf{p}_{b},\mathbf{h}\rangle\right)^{2}$
and (\ref{eq_E_g_e}) for $\mathcal{E}(\mathbf{h},\mathbf{h})$, we
obtain:
\[
\left(\langle\mathbf{p}_{a},\mathbf{h}\rangle-\langle\mathbf{p}_{b},\mathbf{h}\rangle\right)^{2}\le\mathcal{R}^{*}_{ab}\cdot Q(\mathbf{g})\cdot(1+o(1))\le\mathcal{R}^{*}_{ab}\cdot\exp(\beta\rho)\cdot\mathcal{E}(\mathbf{h},\mathbf{h})\cdot(1+o(1)).
\]
Since this inequality holds for any admissible $\mathbf{h}$, the
maximum in Lemma 1 satisfies the estimate:
\[
D^{2}_{\beta}\le\frac{\bigl(\langle\mathbf{p}_{a},\mathbf{h}\rangle-\langle\mathbf{p}_{b},\mathbf{h}\rangle\bigr)^{2}}{\mathcal{E}(\mathbf{h},\mathbf{h})}\le\exp(\beta\rho)\left[\mathcal{R}^{*}_{ab}+O\!\left(e^{-\beta\delta}\right)\right].
\]

Step 3 (completion of the proof). From the lower bound
(\ref{eq_lower_bound_precise}) of Step 1 and the upper bound of Step 2,
the exact asymptotic expansion follows:
\[
D^{2}_{\beta}(\mathcal{B}_{a},\mathcal{B}_{b})=\exp(\beta\rho)\left[\mathcal{R}^{*}_{ab}+O\!\left(e^{-\beta\delta}\right)\right],
\]
where the constant $\mathcal{R}^{*}_{ab}>0$ is determined by formula
(\ref{eq_R_star_def}) and depends exclusively on the local topology of
the minimax edges and the structure of the connected components of
$G_{<\rho}$. Taking the logarithm of both sides, we obtain the classical
logarithmic asymptotics:
\[
\frac{1}{\beta}\ln D^{2}_{\beta}=\rho+\frac{1}{\beta}\ln\left[\mathcal{R}^{*}_{ab}+O\!\left(e^{-\beta\delta}\right)\right]\to\rho\;\text{as }\beta\to\infty,
\]
whence, by virtue of $D_{\beta}>0$ and
$\ln D_{\beta}=\frac{1}{2}\ln D^{2}_{\beta}$, we finally obtain:
\[
\lim_{\beta\to\infty}\frac{1}{\beta}\ln D_{\beta}(\mathcal{B}_{a},\mathcal{B}_{b})=\frac{1}{2}\rho,
\]
which completes the proof of the Theorem.

\section{Discussion}

\label{sec_discussion}

The Theorem formulated in Section \ref{sec_limit_theorem} allows us to
explain the dependence of $u_{\mathrm{nt}}$ on $\Delta F$ observed in
the computational experiment and presented in Table \ref{tab_main}. At
small $\Delta F$ (e.g., $\Delta F=10$ and $20$ kJ/mol), the energy
differences between neighboring vertices are comparable to the thermal
energy $RT$. In this regime, the Kramers matrix is not sharply
heterogeneous: the probabilities of transitions through different edges
differ by no more than $\exp(8)\sim10^{3}$ times. Although this already
creates a certain hierarchy, it is not absolutely rigid. In particular,
alternative paths between two basins, passing through different saddle
points, may have comparable probabilities, and the resulting kinetic
distance is determined by a superposition of the contributions of
several paths. Moreover, in this regime, finite size effects and
differences in basin depths give corrections of the same order of
magnitude as the barrier term $\frac{1}{2}\rho^{(0)}$, which
additionally violates the ultrametric relations. The combination of
these factors leads to a reduction in the fraction of nontrivially
ultrametric triples ($\approx42\%$ and $\approx61\%$, respectively).
With increasing $\Delta F$, the ratio of the characteristic energy
difference to $RT$ increases. At $\Delta F=50$ kJ/mol ($\approx20\,RT$),
the difference in transition rates through different edges can reach
$\exp(20)\sim10^{9}$ times, and $u_{\mathrm{nt}}$ increases to
$\approx81\%$. At $\Delta F=200$ kJ/mol ($\approx80\,RT$), the
difference in transition rates through different edges can reach
$\exp(80)\sim10^{35}$ times. In this regime, the probability of
transition through the highest edge on the path exponentially dominates
over the probabilities of all other transitions, and the kinetics begins
to be determined by a single optimal path -- the one that minimizes the
maximum energy on the path. This corresponds to the construction of the
single-linkage metric $\rho^{(0)}$, which by construction is an
ultrametric. Simultaneously, at large $\Delta F$, the characteristic
amplitude of energies significantly exceeds the heights of suboptimal
barriers, so that the corrections associated with alternative paths
become negligibly small compared to the barrier term
$\frac{1}{2}\rho^{(0)}$. In this regime, the kinetic metric
asymptotically approaches the ultrametric. With a further increase of
$\Delta F$ to $500$ and $1000$ kJ/mol ($\approx200\,RT$ and
$\approx400\,RT$), the dominance of the optimal path and the
suppression of contributions from suboptimal paths become even more
pronounced, and $u_{\mathrm{nt}}$ continues to grow monotonically,
reaching $90.6\%$ and $95.9\%$, respectively.

The physical meaning of the observed transition to ultrametricity lies
in the exponential separation of relaxation time scales. When the
characteristic scale of free energy fluctuations $\Delta F$ becomes
much greater than the thermal energy $RT$, the Arrhenius factors
suppress all transition paths except the minimax ones. In this regime,
the kinetic distance between basins is determined exclusively by the
height of the separating minimax barrier, which is mathematically
equivalent to the single-linkage metric. It is important to emphasize
that the limit transition $\Delta F\to\infty$ at fixed temperature $T$
physically describes a regime of an extremely rugged energy landscape,
in which thermal fluctuations become negligible compared to the
amplitude of energy barriers. This is not equivalent to the
thermodynamic limit of zero temperature ($T\to0$), at which the system
freezes in local minima. In our case, the system remains at a fixed
temperature, but the landscape itself becomes so steep that transitions
between basins are practically forbidden, while relaxation within basins
occurs quickly. It is precisely this regime of limiting landscape
ruggedness that is described by the mathematical abstraction
$\beta\to\infty$, used in the proof of the Theorem.

Nonetheless, even at $\Delta F=1000$ kJ/mol, the degree of
ultrametricity does not reach $100\%$, remaining at a level of
$\approx96\%$. At least two reasons can be indicated for this. First,
the persistence of a small fraction of non-ultrametric triples even at
$\Delta F=1000$ kJ/mol may be associated with finite size effects,
numerical errors in diagonalizing a matrix with a huge spread of
eigenvalues, or topological features of specific realizations of the
graph. Second, according to the proved Theorem, the kinetic metric
contains a pre-exponential factor $\mathcal{R}^{*}_{ab}$, which is
determined by the local combinatorial structure of the graph in the
vicinity of the minimax edges. In our model, the free energies $F_{i}$
are generated from a uniform distribution on $(0,\Delta F)$, which is
equivalent to $F_{i}=\Delta F\cdot\tilde{F}_{i}$, where $\tilde{F}_{i}$
are independent random variables from the interval $\left(0,1\right)$.
The dimensionless energies $\tilde{F}_{i}$ do not depend on $\Delta F$.
Since the constant $\mathcal{R}^{*}_{ab}$ is determined exclusively by
the dimensionless energies $\tilde{F}_{i}$ and the graph topology, it
represents a finite positive quantity that depends on the concrete
realization $\{\tilde{F}_{i}\}$ but does not depend on the scale
$\Delta F$. Consequently, the absolute magnitude of the entropic
correction $\frac{1}{2}\ln\mathcal{R}^{*}_{ab}$ remains a constant of
order $O(1)$ for any $\Delta F$. Although the relative contribution of
this correction to the logarithmic metric decreases as
$O(RT/\Delta F)$, at finite values of $\Delta F$ it gives a non-zero
contribution, leading to deviations from exact ultrametricity.

An important role in the observed effect is played by the graph
topology. Sparse Erdős--Rényi graphs with $\langle k\rangle=2$ are
locally tree-like: for almost all vertices, their $r$-neighborhoods do
not contain cycles with probability tending to unity as $V\to\infty$.
On an ideal tree, between any two vertices there exists exactly one
simple path, and the single-linkage metric is automatically an
ultrametric. The presence of rare cycles in the Erdős--Rényi graph
creates alternative paths, which at finite $\Delta F$ can give a
noticeable contribution and reduce the observed degree of
ultrametricity. However, with increasing $\Delta F$, even these
alternative paths are separated by the characteristic heights of
barriers, and their contribution is exponentially suppressed. Thus, the
fundamental source of ultrametricity is precisely the large energy
spread, and not the graph topology. However, the local tree-likeness
of sparse graphs acts as a crucial practical factor: it minimizes the
number of suboptimal paths, thanks to which asymptotic ultrametricity
becomes kinetically manifest already at moderate values of $\Delta F$.
The proved Theorem formally holds for an arbitrary finite connected
graph, including a complete graph. However, at $\langle k\rangle\gg1$,
the number of alternative paths is exponentially large, and
substantially larger values of $\Delta F/(RT)$ would be required to
suppress their contributions, which complicates the direct numerical
observation of the effect.

Based on the obtained results, one can try to establish a connection
between the kinetic metric and the $p$-adic parametrization of
ultrametric spaces, used in \cite{avetisov2002,ABZ_2014,avetisov2009,bikulov2021}
to describe the conformational dynamics of a protein. In these models,
the state space is assumed to be isomorphic to the set of leaves of a
regular $p$-adic tree, and the distance between states is given by an
explicit ultrametric. The limit theorem proved in Section
\ref{sec_limit_theorem} shows that for a graph of arbitrary topology
with an exponentially wide energy distribution, the kinetic metric
asymptotically converges to the single-linkage ultrametric without
a priori postulating a tree-like structure. This connection acquires a
concrete dynamical content if we consider a coarse-grained Markov
dynamics directly on the space of basins of attraction. Using the
kinetic metric as a basis for determining the rates of inter-basin
transitions, one can write an evolution equation of the form
\begin{equation}
\frac{df_{a}(t)}{dt}=\sum_{b\neq a}D^{-2}_{\beta}(\mathcal{B}_{a},\mathcal{B}_{b})\bigl(f_{b}(t)-f_{a}(t)\bigr),\label{eq_basin_dynamics}
\end{equation}
where $f_{a}(t)$ is the probability of finding the system in basin
$\mathcal{B}_{a}$ at time $t$. The matrix of coefficients
$W_{ab}=D^{-2}_{\beta}(\mathcal{B}_{a},\mathcal{B}_{b})$ for $a\neq b$
satisfies all axioms of a continuous-time Markov chain generator. The
proposed equation represents a Markovian closure, which is
mathematically correct only in the limit of infinite separation of time
scales ($\Delta F/RT\to\infty$), when intra-basin relaxation occurs
infinitely fast compared to inter-basin transitions, and the
non-Markovian dynamics reduces asymptotically to a Markov jump process.

Physically, equation (\ref{eq_basin_dynamics}) describes hopping
dynamics on the space of basins. In the regime of high barriers, we
obtain the classical Arrhenius asymptotics:
\begin{equation}
D^{-2}(\mathcal{B}_{a},\mathcal{B}_{b})\;\sim\;\frac{1}{\mathcal{R}^{*}_{ab}}\,\exp\!\left(-\frac{\rho^{(0)}_{\mathrm{phys}}(\mathcal{B}_{a},\mathcal{B}_{b})}{RT}\right),\label{eq_rate_asympt}
\end{equation}
where the pre-exponential factor $(\mathcal{R}^{*}_{ab})^{-1}$ depends
on the pair of basins and is determined by the local topology of the
minimax edges. Thus, the proposed dynamics asymptotically turns into the
classical Arrhenius jump dynamics with physical minimax barriers.

The structural analogy of equation (\ref{eq_basin_dynamics}) with
$p$-adic diffusion models is deep and substantive. In
\cite{avetisov2002,avetisov2009,ABZ_2014,bikulov2021}, the dynamics of
conformational states of a protein is described by a kinetic equation on
a $p$-adic space, in which the evolution operator has the form of an
integral operator with a kernel depending on the ultrametric distance.
The hierarchical block structure of the matrix $D^{-2}_{\beta}$,
arising from the approximate ultrametricity of $D_{\beta}$, reproduces
the hierarchy of relaxation times characteristic of $p$-adic models. The
principal difference of the present approach from $p$-adic models is
that the ultrametric structure is not postulated a priori, but arises as
an asymptotic property of the kinetic metric computed from first
principles based on the microscopic Kramers matrix.

It is necessary to point out a substantial limitation of equation
(\ref{eq_basin_dynamics}) as a physical model. Since the kinetic metric
$D$ is symmetric, the rate matrix $W_{ab}$ is also symmetric, and the
stationary distribution of the process is uniform over all basins. This
contradicts the thermodynamic requirement that in the original
microscopic Kramers dynamics, the stationary probability of being in a
basin should be proportional to its statistical weight. To restore
thermodynamic correctness, one can modify the equation by introducing
factors ensuring the fulfillment of the detailed balance condition:
\begin{equation}
\widetilde{W}_{ab}=\sqrt{\frac{W_{a}}{W_{b}}}\,D^{-2}_{\beta}(\mathcal{B}_{a},\mathcal{B}_{b}),\label{eq_detailed_balance}
\end{equation}
which leads to the modified evolution equation
\begin{equation}
\frac{df_{a}(t)}{dt}=\sum_{b\neq a}\sqrt{\frac{W_{a}}{W_{b}}}\,D^{-2}_{\beta}(\mathcal{B}_{a},\mathcal{B}_{b})\,f_{b}(t)-f_{a}(t)\sum_{b\neq a}\sqrt{\frac{W_{b}}{W_{a}}}\,D^{-2}_{\beta}(\mathcal{B}_{a},\mathcal{B}_{b}),\label{eq_basin_dynamics_db}
\end{equation}
whose stationary distribution is the Boltzmann distribution
$\pi_{a}=W_{a}/Z$. The factor $\sqrt{W_{a}/W_{b}}$ does not violate
the ultrametric hierarchy of rates, since its logarithmic contribution
is of order $O(1)$ compared to the leading barrier term
$O(\Delta F/RT)$.

The totality of these observations indicates that $p$-adic diffusion
equations may be applicable to systems with strong energy heterogeneity,
going beyond the biopolymer applications considered in
\cite{avetisov2002,ABZ_2014,avetisov2009,bikulov2021}. The present
work shows that for such systems, the ultrametric organization and the
corresponding hierarchical dynamics arise as a universal asymptotic
property, requiring neither biological optimization nor an a priori
specification of a tree-like structure of the state space.

The obtained results also define a number of directions for further
research. First, it is of interest to quantitatively estimate the
contribution of suboptimal paths to the magnitude of $u_{\mathrm{non}}$
at finite $\Delta F$ and its dependence on the parameters $\Delta F$,
$T$, and the average degree of the graph $\langle k\rangle$. Second,
the model used in this work assumes a uniform distribution of the free
energies of the vertices. Considering models with other distributions
will allow establishing whether the type of distribution influences the
rate of convergence of $u_{\mathrm{nt}}$ to unity with increasing
$\Delta F$. Third, it is of interest to apply the developed formalism to
random graphs with a higher average degree ($\langle k\rangle>2$), as
well as to graphs containing a pronounced modular structure, which will
allow assessing the influence of the number of cycles and topology
heterogeneity on the degree of ultrametricity at fixed parameters of the
energy landscape. Fourth, the coarse-grained dynamics
(\ref{eq_basin_dynamics_db}) proposed in this section deserves an
independent investigation. It is of interest to compare its relaxation
properties (spectrum of relaxation times, autocorrelation functions)
with the results of direct modeling of the microscopic Kramers dynamics
on the graph, as well as to establish exact conditions under which the
replacement of the microscopic generator $K$ by an effective ultrametric
operator on the space of basins is asymptotically exact.

\section*{Conflict of Interest}

The author declares no conflict of interest.

\section*{Funding}

The study was carried out without external funding sources.

\section*{Data and Code Availability}

The source code of the program is openly available in the Zenodo repository
\cite{zubarev2026_github}. Additional data supporting the findings
of this study are available from the author upon reasonable request.

\end{document}